\documentclass[reprint,amsmath,amssymb,aps,prb,longbibliography]{revtex4-1}

\usepackage{graphicx}
\usepackage{color}
\usepackage{amsmath}
\usepackage{dcolumn}
\usepackage{enumitem}
\usepackage{epstopdf}

\begin{document}

\title{Interplay of magnetization dynamics with a microwave waveguide at cryogenic temperatures}

\author{I.~A.~Golovchanskiy$^{1,2}$, N.~N.~Abramov$^{2}$, M.~Pfirrmann$^{3}$, T.~Piskor$^{3}$, J.~N.~Voss$^{3}$, D.~S.~Baranov$^{1,4,5}$, R.~A.~Hovhannisyan$^{1}$, V.~S.~Stolyarov$^{1,4,6}$, C.~Dubs$^{7}$, A.~A.~Golubov$^{1,8}$, V.~V.~Ryazanov$^{2,4,6}$, A.~V.~Ustinov$^{2,3}$, M.~Weides$^{3,9}$}

\affiliation{
$^{1}$ Moscow Institute of Physics and Technology, State University, 9 Institutskiy per., Dolgoprudny, Moscow Region, 141700, Russia \\
$^{2}$ National University of Science and Technology MISIS, 4 Leninsky prosp., Moscow, 119049, Russia \\
$^{3}$ Physikalisches Institut, Karlsruhe Institute of Technology, 76131 Karlsruhe, Germany \\
$^{4}$ Institute of Solid State Physics (ISSP RAS), Chernogolovka, Moscow Region 142432, Russia \\
$^{5}$ Laboratoire de Physique et d’Etude des Materiaux, UMR8213, {\'E}cole sup{\'e}rieure de physique et de chimie industrielles de la Ville de Paris, Paris Sciences et Lettres Research University, Institut des NanoSciences de Paris-Sorbonne Universite, 10 rue Vauquelin, 75005 Paris, France \\
$^{6}$ Solid State Physics Department, Kazan Federal University, Kazan 420008, Russia \\
$^{7}$ INNOVENT e.V. Technologieentwicklung, Pruessingstrasse 27B, 07745 Jena, Germany \\
$^{8}$ Faculty of Science and Technology and MESA+ Institute for Nanotechnology, University of Twente, Enschede 7500 AE, The Netherlands \\
$^{9}$ School of Engineering, University of Glasgow, Rankine Building, Oakfield Avenue, Glasgow G12 8LT, United Kingdom
}%

\begin{abstract}
In this work, magnetization dynamics is studied at low temperatures in a hybrid system that consists of a thin epitaxial magnetic film coupled with a superconducting planar microwave waveguide.
The resonance spectrum was observed over a wide magnetic field range, including low fields below the saturation magnetization and both polarities.
Analysis of the spectrum via a developed fitting routine allowed for the derivation of all magnetic parameters of the film at cryogenic temperatures, the detection of waveguide-induced uniaxial magnetic anisotropies of the first and the second order, and the uncovering of a minor misalignment of magnetic field.
A substantial influence of the superconducting critical state on resonance spectrum is observed and discussed.
\end{abstract}

\maketitle

\section{Introduction}

The field of magnonics studies the application of magnetization oscillations and waves in ferromagnetic structures \cite{Lenk_PR_507_107,Chumak_NatPhys_11_453,Kajiwara_Nat_464_262,Spec_Iss,Demokritov_book,Evelt_PRAppl_10_041002}.
The following benefits make magnonics promising for application in processing of microwave signals:
tunability of the magnon dispersion with applied magnetic field and the geometry of the medium,
low dissipation and power consumption, high operational frequencies,
convenient micron and sub-micron scales of spin wavelength at microwave frequencies,
and, finally, absence of parasitic coupling of spin waves with nonmagnetic environments.
Conventionally, magnonics is a room-temperature research discipline.

Currently a sub-discipline is emerging that deals with magnetization dynamics at cryogenic temperatures and can be referred to as ``cryogenic magnonics''.
Indeed, quantum magnonics is of high current interest \cite{Huebl_PRL_111_127003,Tabuchi_PRL_113_083603,Zhang_PRL_113_156401,Morris_SciRep_7_11511,Weides}.
Microwave experiments in quantum magnonics are typically performed at milli-kelvin temperatures, often using setups equipped with superconducting quantum circuits.
On the other hand, a development of various hybrid devices is taking place based on superconducting resonators\cite{Huebl_PRL_111_127003,Golovchanskiy_JAP_123_173904} and Josephson junctions  \cite{Barnes_SUST_24_024020,Mai_PRB_84_144519,Golovchanskiy_SUST_30_054005}.
Also, it was shown that hybridization of a magnon medium with superconducting structures results in substantial modification of dispersion properties\cite{Lebed_PisZTF_15_27,Anfinogenov_PisZTF_15_24,Golovchanskiy_AFM_28_1802375,Golovchanskiy_PRB_sub}, as well as in the formation of magnonic band structures\cite{Dobrovolskiy_NatPhys}. 
Lastly, metamaterial properties have been reported for superconductor/ferromagnet superlattices \cite{Pimenov_PRL_95_247009}.
More generally and beyond superconductor-induced phenomena, the magnetic properties at low temperatures are probed in absence or only with minor thermal excitations. 
Typical thermal effects for standard magnonics, such as reduced saturation magnetization or thermally activated domain wall motion, are lessened for cryo-magnonics, leading to new phenomena in ferromagnetic resonance (FMR).

In this regard, investigation of magnetic properties of ferromagnetic films at low temperature as well as of their interaction with superconducting circuits is imperative.
This report addresses both problems.
We focus on the ferromagnetic resonance in a thin Yttrium Iron Garnet (YIG) film coupled to a superconducting Nb planar waveguide in out-of-plane magnetic field.
We obtain the FMR spectrum at low temperature in a wide field range. 
The spectrum shows linear magnetic resonance versus field dependence for high fields and a range of nonlinear dependence of FMR frequency at low magnetic fields where the Kittel formulas are inapplicable.
Developing a fitting routine, we derive all magnetic parameters of the YIG film.
Our analysis shows that the waveguide itself induces substantial uniaxial magnetic anisotropy.
Next, we study the FMR spectrum at temperatures below the superconducting critical temperature of the waveguide and observe an influence of the superconducting critical state of Nb on the resonance spectrum.

We note that while YIG is probably the most popular magnetic material for magnonic applications, owing to its low damping, 
the damping in YIG and its temperature dependence are not addressed in this paper and can be found elsewhere \cite{Haidar_JAP_117_17D119,Beaulieu_IEEEMagLett_9_3706005,Boventer_PRB_97_184420}. 
In this report, YIG is selected as a model magnetic single-crystalline thin film with distinct magnetocrystalline anisotropy and sufficiently low saturation magnetization, which is convenient for out-of-plane measurements.

\section{Experimental details}

%
\begin{figure}[!ht]
\begin{center}
\includegraphics[width=0.95\columnwidth]{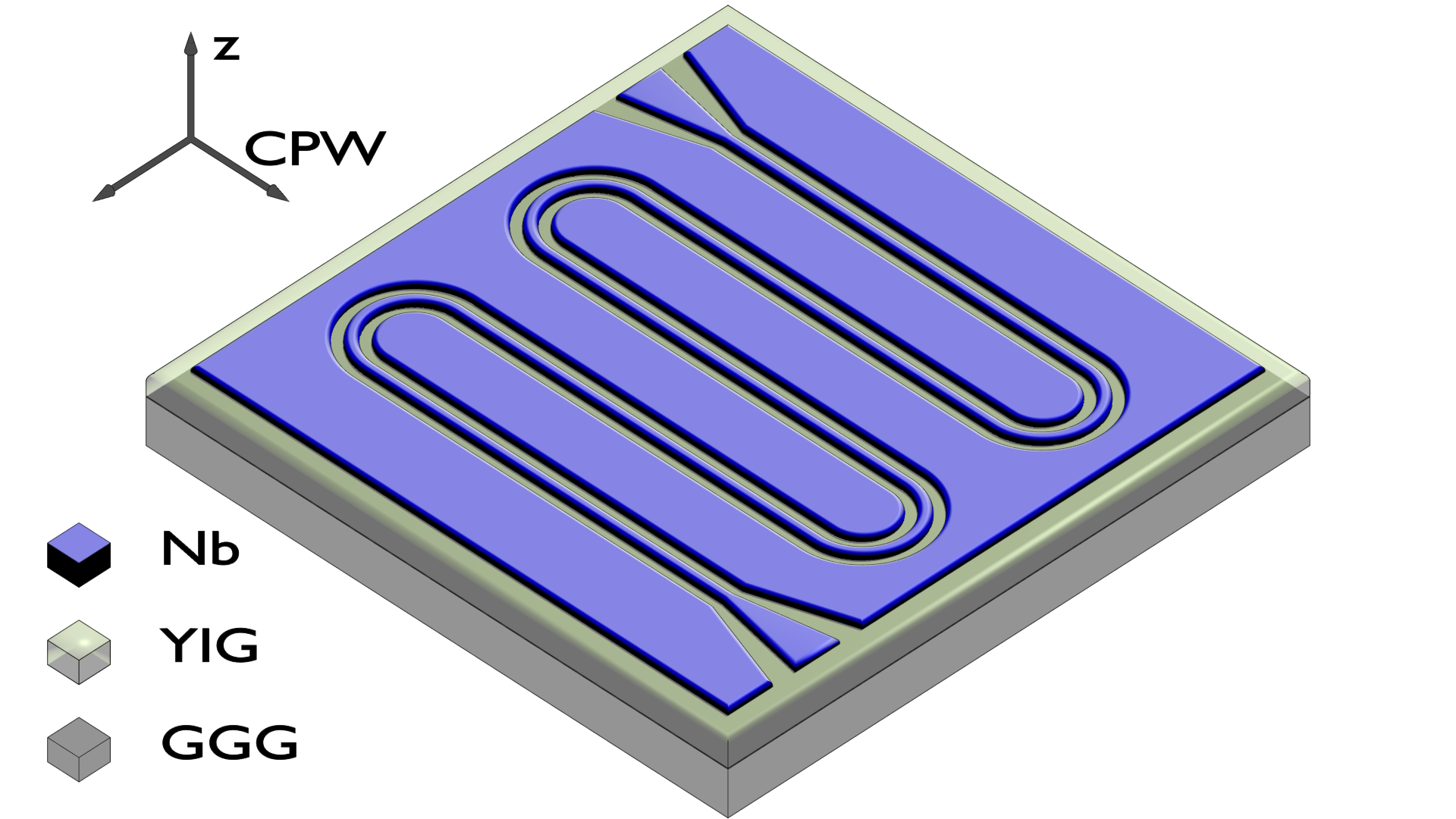}
\caption{
Schematic illustration of the investigated system.  
YIG epitaxial film is grown on [111]-oriented single-crystal GGG substrate by means of LPE.
A Nb CPW is fabricated directly on top of YIG film.
The main direction of the CPW is indicated with the axis.
Magnetic field $H$ is applied out-of-plane, along [111] orientation of YIG/GGG (i.e., along $z$-direction). 
}
\label{Sam}
\end{center}
\end{figure}

The FMR absorption measurements were performed using the so-called VNA-FMR approach \cite{Neudecker_JMMM_307_148,Kalarickal_JAP_99_093909,Chen_JAP_101_09C104} (VNA stands for the vector network analyzer).
A schematic illustration of the investigated system is shown in Fig.~\ref{Sam}.
The single-crystalline epitaxial YIG film of thickness $d=51$~nm was deposited on single-crystal [111]-oriented Gadolinium Gallium Garnet (GGG) substrate using the liquid phase epitaxy (LPE) technique.
Details of LPE as well as room-temperature characteristics of LPE-grown ultra-thin YIG films can be found elsewhere \cite{Dubs_JPDAP_50_204005,Beaulieu_IEEEMagLett_9_3706005}.
Measurement of the FMR response in YIG was enabled by fabrication of a co-planar waveguide (CPW) directly on top of YIG film.
The CPW was patterned out of 150~nm thick magnetron-sputtered niobium (Nb) thin film with superconducting critical temperature $T_c\simeq8.5$~K using photo-lithography and plasma-chemical etching.
Deposition of Nb at room temperature was obstructed by poor adhesion of the metal film to the YIG surface, and therefore, was performed at 300$^{\circ}$C.
The 50~$\Omega$ impedance of the superconducting CPW was provided by its 27-40-27 $\mu$m gap-center-gap dimensions.
Direct placement of the CPW on the probed magnetic film and its elongation via meandering enhance sensitivity to weak FMR absorptions\cite{Golovchanskiy_JAP}.
The experimental chip was installed in a copper sample holder and wire bonded to a PCB with RF connectors. 
A thermometer and a heater were attached directly to the holder for precise temperature control.
The holder was placed in a superconducting solenoid inside a closed-cycle cryostat (Oxford Instruments Triton, base temperature 1.2 K).
The response of the system was studied by analyzing the transmitted microwave signal S$_{21}$ with the VNA Rohde \& Schwarz ZVB20.

\section{Results and discussion}

\begin{figure*}[!ht]
\begin{center}
\includegraphics[width=1\columnwidth]{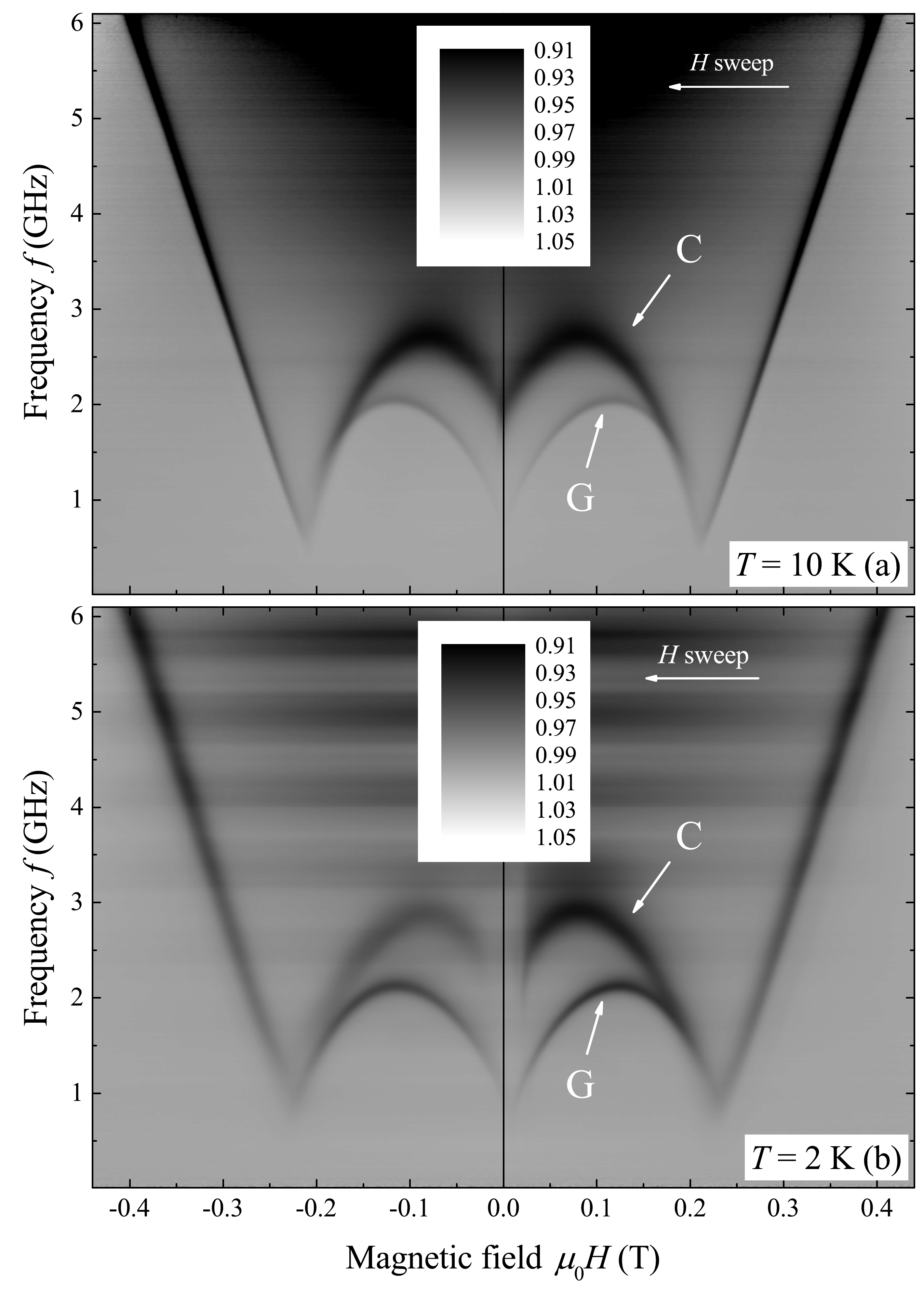}
\includegraphics[width=1\columnwidth]{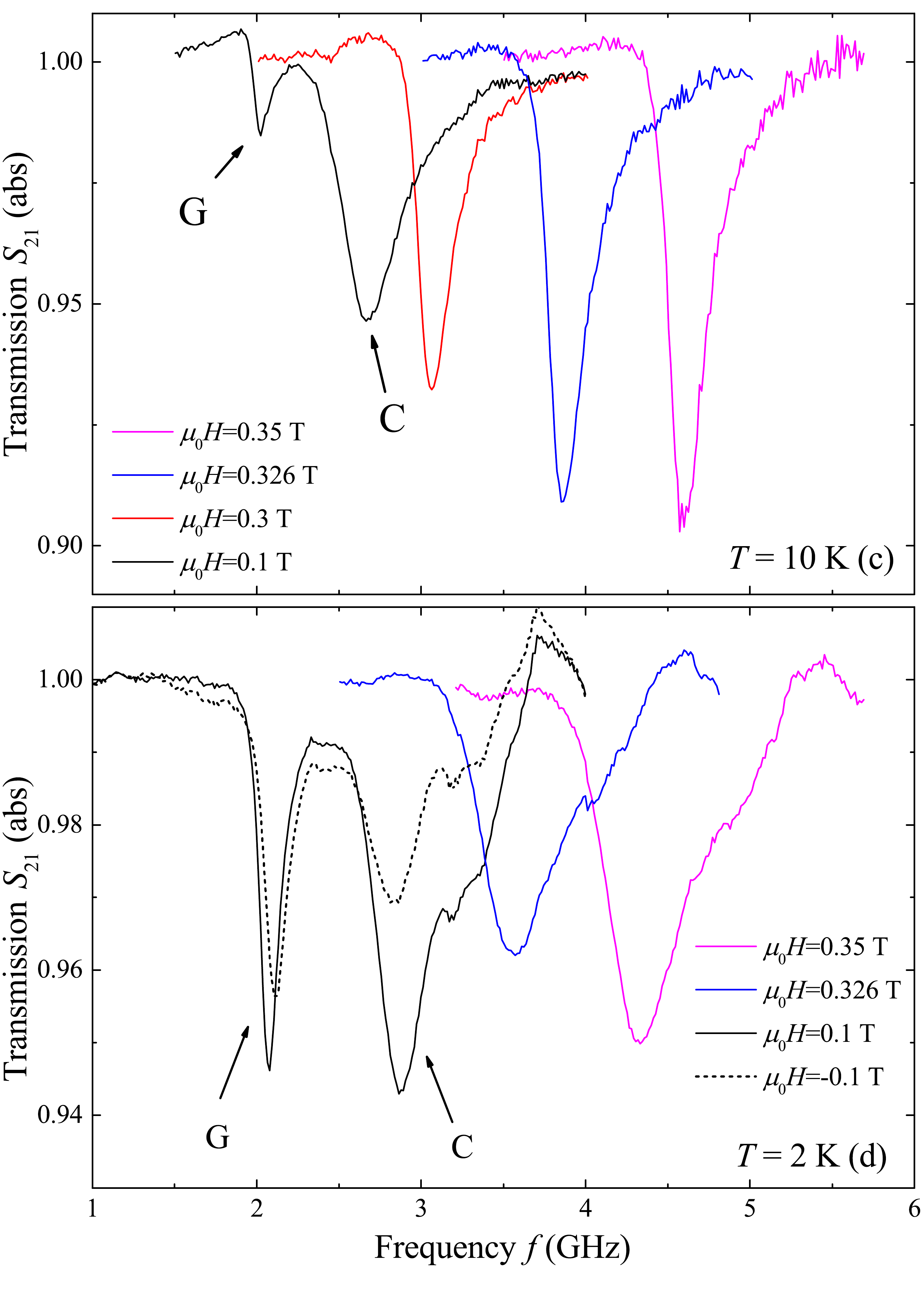}
\caption{a) and b) Gray-scale-coded transmission spectra $|S_{21}(\mu_0H,f)/S_{21}(\mu_0 H=0.5$~T$,f)|$ measured at $T=10$~K above $T_{\rm c}$ (a) and $T=2$~K below $T_{\rm c}$ (b).
c) and d) Corresponding frequency dependencies of the normalized transmission $|S_{21}(f)|$ at several magnetic fields at $T=10$~K (c) and $T=2$~K (d).
For curves in (c) and (d) the background was subtracted.
At $T<T_{\rm c}$ the spectrum shows hysteresis of absorption.
Magnetic field was swept negatively from +0.5~to~-0.5~T (indicated with arrows), and, therefore, the part of spectra in (b) at positive fields provides the ``down-field-sweep'' data while the part of the spectrum at negative fields provides the ``up-field-sweep'' data.
Labels C and G indicate higher- and lower-frequency spectral lines, respectively.}
\label{raw}
\end{center}
\end{figure*}

Figures~\ref{raw}a,b show transmission spectra of the studied sample at $T=10$~K$>T_{\rm c}$ of Nb and at 2~K$<T_{\rm c}$.
Spectra have been normalized with $S_{21}(f)$ at $\mu_0 H=0.5$~T.
Figures~\ref{raw}c,d show a set of normalized absorption curves $S_{21}(f)$ of the sample at the same temperatures and several magnetic fields.
Field dependent spectral lines in Figs.~\ref{raw}a,b with the minimum transmission correspond to FMR curves $f_{\rm r}(\mu_0 H)$.
Both spectra show linear FMR response at $|\mu_0 H|>0.2$~T, which is typical for the Kittel-FMR mode of a thin film in out-of-plane magnetic field.
The resonance frequency with out-of-plane field is $f_{\rm r}\propto(\mu_0 H-4\pi M_{\rm eff})$, which indicates the value of the effective saturation magnetization $4\pi M_{\rm eff}\sim2000$~Oe at $f_{\rm r}\rightarrow 0$.
Upon decreasing $|\mu_0 H|$, the linear resonance line is terminated with a kink at $|\mu_0 H|\sim 4\pi M_{\rm eff}$ and transforms into two FMR branches with nonlinear dependence of resonance frequency versus field for $|\mu_0 H|<4\pi M_{\rm eff}$.
We refer to the higher-frequency FMR branch with stronger absorption as C-line and to the lower-frequency FMR branch with weaker absorption as G-line.
Note that in general, observation of FMR in thin films with out-of-plane geometry at $|\mu_0 H|<4\pi M_{\rm eff}$ might be challenging due to formation of nonuniform magnetization configurations.
With out-of-plane field $|\mu_0 H|<4\pi M_{\rm eff}$, ferromagnetic films are not magnetized to saturation, and Kittel formulas for FMR are not applicable.
Splitting of the FMR response into several spectral lines at $|\mu_0 H|<4\pi M_{\rm eff}$ can be caused by various factors, including standing spin wave resonances \cite{Kittel_PR_73_155,Khivintsev_JAP_108_023907,Klingler_JPDAP_48_015001,Golovchanskiy_AFM_28_1802375}, FMR response of magnetic domain structure \cite{Artman_JAP_49_1587,RAMESH_JMMM_74_123,Camara_JPCM_29_465803}, or magnetic phase separation.

After transition of the Nb CPW into the superconducting state, the transmission spectrum changes (compare Figs.~\ref{raw}a and b, Figs.~\ref{raw}c and d).
While the spectrum at $T<T_{\rm c}$ consists of the same resonance lines as at $T>T_{\rm c}$, superconductivity manifests itself in hysteresis of FMR peak absorption at $|\mu_0 H|<0.2$~T, which is best visible for the C-line (compare $S_{21}(f)$ at $\mu_0 H=0.1$~T and $\mu_0 H=-0.1$~T in Fig.~\ref{raw}d):
FMR absorption at negatively swept magnetic field (positive $H$ in Figs.~\ref{raw}b,d) is substantially stronger than at positively swept magnetic field (negative $H$ in Figs.~\ref{raw}b,d).
In addition, at $T<T_{\rm c}$ a suppression of FMR response is observed at the low field region $|\mu_0 H|<0.02$~T.
Below we will discuss the FMR response of YIG in absence of superconductivity, establish causes for the split of FMR at $|\mu_0 H|<4\pi M_{\rm eff}$, and define the contribution of superconductivity to the FMR spectrum.

\subsection{FMR at $T>T_{\rm c}$. Magnetic properties of YIG film at cryogenic temperatures.}

Having analysed possible origins for the split of the FMR into the C-line and G-line in Fig.~\ref{raw} we can state that neither domain structure nor spin waves can contribute to the FMR spectrum for our particular study. 
For instance, nucleation of magnetic domains upon demagnetization at $\mu_0 H<4\pi M_{\rm eff}$ occurs for thin films with strong perpendicular anisotropy in comparison with the demagnetizing energy \cite{BLAKE_IEEE_18_985,Virot_JPDAP_45_405003}, i.e. when the magnetic quality parameter $Q=K_{\rm u}/2\pi M_{\rm s}^2>1$, where $K_{\rm u}$ is the out-of-plane uniaxial anisotropy, and $M_{\rm s}$ is the saturation magnetization.
However, a typical field of uniaxial anisotropy $\mu_0 H_{K_{\rm u}}=2K_{\rm u}/M_{\rm s}$ in LPE-grown YIG thin films ranges up to $\sim200$~Oe \cite{Dubs_JPDAP_50_204005,Klingler_JPDAP_48_015001,Lee_JAP_120_033905,Beaulieu_IEEEMagLett_9_3706005} ensuring $Q\ll 1$.
The highest values of uniaxial anisotropy in YIG films $\mu_0 H_{K_{\rm u}}\sim10^3$~Oe that can be obtained in pulsed-laser-deposited films \cite{Manuilov_JAP_106_123917} still ensure $Q<1$.
As an additional test, we have performed a magnetic force microscopy study of magnetic flux structure at the surface of the YIG film at 4~K using attocube attoDRY 1000 closed-cycle cryogenic microscope, supplied with a superconducting solenoid, and found no traces of domains or any field-dependent magnetic structure. 
Therefore, we confirm that formation of the domain structure does not occur.
The magnetic state of the YIG film is single-domain, and variation of the out-of-plane component of magnetization at $\mu_0 H<4\pi M_{\rm eff}$ occurs via rotation of the magnetization vector from out-of-plane orientation to in-plane.

The absence of contribution of standing spin-wave resonances to the FMR spectrum can be illustrated in the following way.
At $H=0$ the magnetization vector of a single-domain film is aligned in-plane.
Therefore, the Kittel formula for FMR and dispersion relations for any spin-wave mode at $H=0$ become applicable.
When several resonances are observed, a contributing spin-wave mode can be identified by estimating a resonance frequency difference $\Delta f_{\rm r}$ between the Kittel mode and any standing spin-wave resonance mode.
The latter appears due to quantization of the wavelength with geometrical parameters of a sample.
The difference $\Delta f_{\rm r}$ is then compared with the experimentally observed one $\sim1.3$~GHz at $H=0$ (Fig.~\ref{raw}).
If an in-plane magnetostatic standing spin-wave mode\cite{Stancil,Serga_JPDAP_43_264002} is assumed, e.g., the backward volume mode or the magnetostatic surface mode, its wavelength $\lambda/2$ for the standing mode should be quantized with dimensions of the CPW, i.e., $\lambda/2\sim20-40$~$\mu$m.
Such a standing spin-wave mode provides only marginal difference $\Delta f_{\rm r}\lesssim10$~MHz due to a small ratio $d/\lambda\sim10^{-3}$.
Alternatively, if exchange-dominated perpendicular standing spin-wave resonance\cite{Kittel_PR_100_1295,Seavey_JAP_30_S227} is assumed the typical exchange constant in YIG films\cite{Klingler_JPDAP_48_015001} $\sim4\times10^{-12}$~J/m provides $\Delta f_{\rm r}\sim2.5$~GHz for $d=\lambda/4$ and $\Delta f_{\rm r}\sim7.5$~GHz for $d=\lambda/2$.
Thus, none of the possible standing spin-wave modes can provide $\Delta f_{\rm r}\sim1.3$~GHz.
Overall, when a standing spin-wave resonance is excited multiple consequential spectral lines are expected.
FMR absorption for each line should decrease progressively with the mode number (see, for instance Ref.~\cite{Bunyaev_SRep_5_18480}).
Such a picture is not observed in our experiment.
Therefore, we confirm that several spectral lines in Fig.~\ref{raw} at $|\mu_0 H|<4\pi M_{\rm s}$ are not caused by standing spin-wave resonances.

The remaining explanation for two FMR lines requires the existence of two resonating areas with different magnetic properties in the vicinity of the CPW.
The magnetic structure is essentially single-domain in each area.
The resonating areas can be identified by the coupling strength of microwaves to precessing magnetization that is proportional to the FMR amplitude and correlates directly with the amplitude of excitation AC magnetic fields.
In CPW geometry, AC magnetic fields are mainly focused in the vicinity of the central transmission line\cite{Neudecker_JMMM_307_148,Golovchanskiy_JAP_123_173904}.
Therefore, geometry of the experiment (Fig.~\ref{Sam}) suggests that the lower-frequency, weaker G-line originates from YIG at gap areas of CPW where the coupling is weaker, while higher-frequency, stronger C-line appears due to FMR response of YIG area under the central conducting line of the CPW.
The accuracy of that explanation is strengthened by additional features, as discussed below. 

\begin{figure}[!ht]
\begin{center}
\includegraphics[width=0.95\columnwidth]{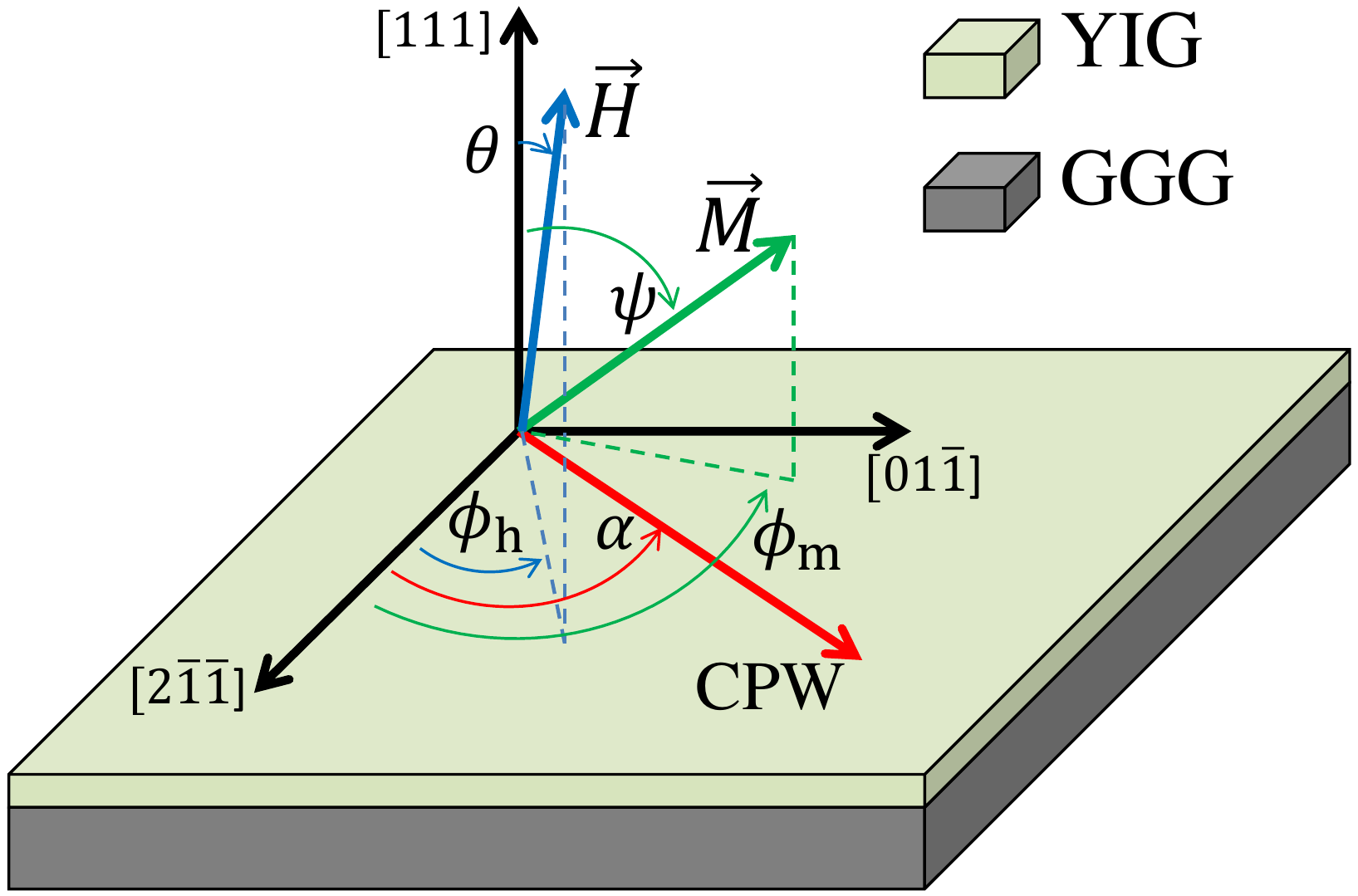}
\caption{
Spherical coordinate system for the studied YIG film sample.
Adopted from Ref.~\cite{Lee_JAP_120_033905}.
Direction of the CPW transmission line (see Fig.~\ref{Sam}) is indicated with the red axis,
which specifies additional in-plane uniaxial anisotropy due to CPW directionality .
}
\label{Coord}
\end{center}
\end{figure}

For the case of the single-domain single-crystalline YIG film, the analytical resonance curve $f_{\rm r}(\mu_0 H)$ can be obtained in the entire $H$-range following Refs.~\cite{Lee_JAP_120_033905,SMIT_PRR_10_113,SUHL_PR_97_555,Rezende_PRB_49_15105} (we keep the notations given in Ref.~\cite{Lee_JAP_120_033905}).
The orientation of magnetization of a single-domain film at arbitrarily oriented magnetic field is defined by the minimum of free magnetostatic energy $g=g(M_{\rm s},h,k_1,k_2,k_{\rm u},\theta,\psi,\phi_{\rm h},\phi_{\rm m})$, where $k_1$, $k_2$ and $k_{\rm u}$ are unitless parameters of cubic magnetocrystalline anisotropy and out-of-plane uniaxial anisotropy, respectively, 
$h=\mu_0 H/4\pi M_{\rm s}$ is the normalized external magnetic field, and $\theta,\psi,\phi_{\rm h},\phi_{\rm m}$ define the orientations of $H$ and $M_{\rm s}$ with respect to principle crystallographic axes of YIG in spherical coordinates (see Fig.~\ref{Coord}).
In addition, the system in Fig.~\ref{Sam} has a distinct directionality along the orientation of the CPW.
This directionality may contribute to the orientation of magnetization.
We account for its possible contribution by an additional energy term $g_{\rm a}$ added to the free magnetostatic energy $g$ that provides a phenomenological in-plane uniaxial anisotropy of the first order. 
The term of the in-plane uniaxial anisotropy of the first order in the coordinates of Fig.~\ref{Coord} is 
\begin{equation}
g_{\rm a}=-k_{\rm a1}\sin(\psi)^2\cos(\phi_{\rm h}-\alpha)^2.
\label{g_a1}
\end{equation}
FMR frequency is defined by derivatives at position of minimum of free energy\cite{SMIT_PRR_10_113,SUHL_PR_97_555,Rezende_PRB_49_15105} as
\begin{equation}
f_{\rm r}\sim \gamma(g_{\psi\psi}g_{\phi_{\rm m}\phi_{\rm m}}-g_{\psi\phi_{\rm m}}^2)^{1/2}/\sin(\psi),
\label{f_r1}
\end{equation}
where $\gamma$ is the gyromagnetic ratio.
See Refs.~\cite{SMIT_PRR_10_113,SUHL_PR_97_555,Rezende_PRB_49_15105,Lee_JAP_120_033905} for details.

\begin{figure}[!ht]
\begin{center}
\includegraphics[width=0.95\columnwidth]{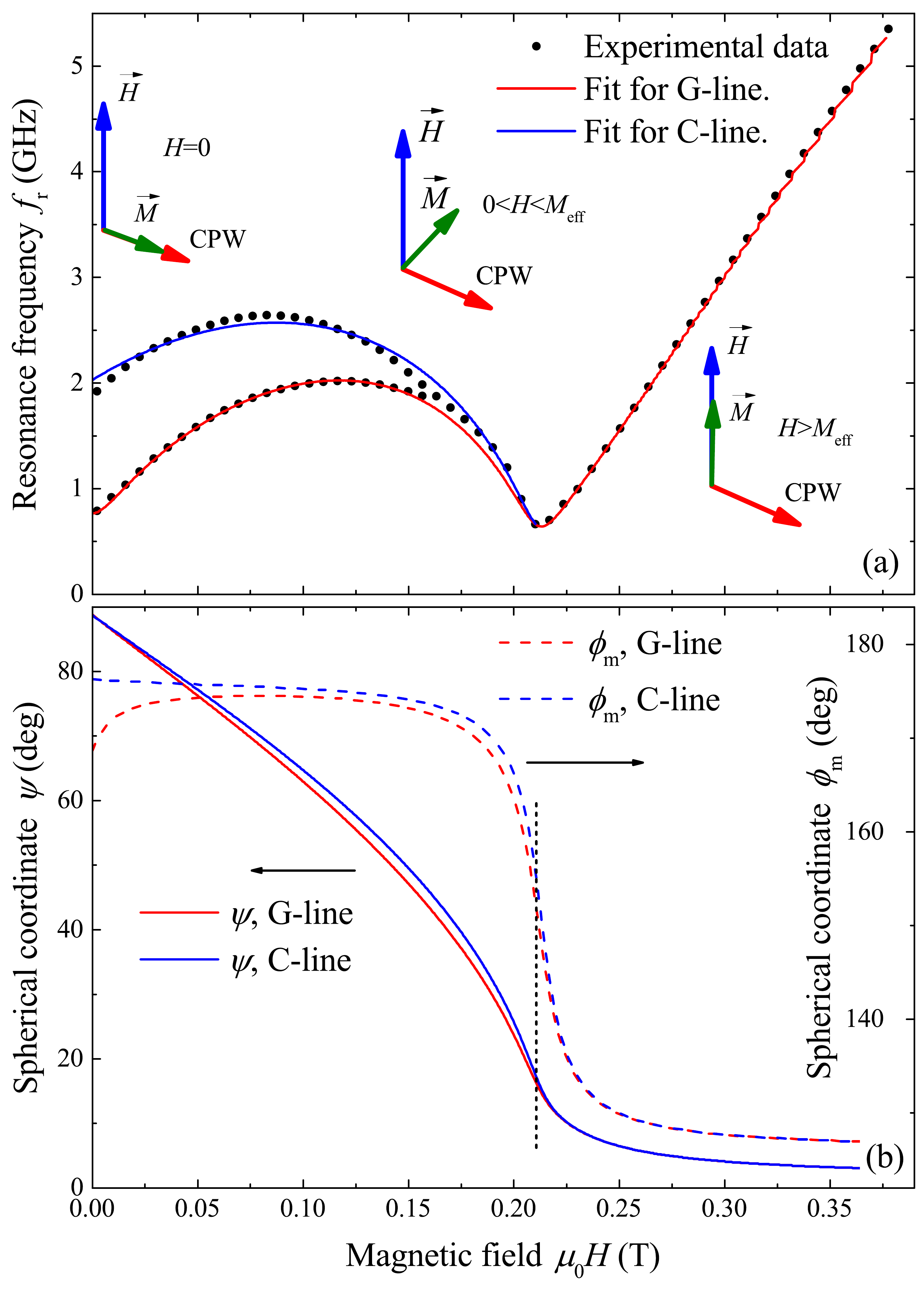}
\caption{Fitting experimental $f_r(\mu_0 H)$.
(a) The dependence of the FMR frequencies on magnetic field. 
The CPW induces the uniaxial anisotropy of the 1st (G-line) and the 2nd order (C-line).
Pictograms in (a) illustrate orientation of magnetization relative to orientations of magnetic field and CPW axis at different magnetic fields.
(b) The dependence of free-energy-minimum orientations of magnetization on magnetic field.
The black dotted line indicates position of the kink in FMR curves.}
\label{fit}
\end{center}
\end{figure}

Dotted data in Fig.~\ref{fit}a show the experimental $f_{\rm r}(\mu_0 H)$ resonance curves extracted from Fig.~\ref{raw}a.
First, we focus on the G-line of the FMR spectrum.
In order to fit the data, we have developed the following routine, which allowed us to obtain all magnetic parameters of $g$ and $f_{\rm r}$, despite a large number of parameters and their partial interdependency.
First, we note that when misalignment $\theta$ of orientation of magnetic field with the $z$-axis is small, and in-plane CPW uniaxial anisotropy $k_{\rm a1}$ is negligible, the linear part at $\mu_0 H\gg 4\pi M_{\rm eff}$ can be fitted with the simplified expression\cite{Lee_JAP_120_033905} 
\begin{equation}
f_{\rm r}=2\gamma M_{\rm s}(h-(1-k_{\rm u}+2k_1/3+2k_2/9)).
\label{f_r2}
\end{equation}
Fitting the data in the field range from 0.3~T to 0.4~T with Eq.~\ref{f_r2}, we obtain the gyromagnetic ratio $\gamma/2\pi=2.985$~GHz/kOe, which is close to the ratio for a free electron $2.803$~GHz/kOe, and also the value of saturation magnetization in relation with anisotropy parameters $4\pi M_{\rm eff}=4\pi M_{\rm s}\times(1-k_{\rm u}+2k_1/3+2k_2/9)=1975$~Oe.
Next we note that 
(i) the position of the kink at $\mu_0 H\simeq0.22$~T in the $(f_{\rm r},\mu_0 H)$ plot is mostly determined by misalignment of the magnetic field with $z$-axis, i.e., by $\theta$ and $\phi_{\rm h}$;
(ii) the position of the maximum of FMR frequency at $\mu_0 H\simeq0.12$~T in the $(f_{\rm r},\mu_0 H)$ plot is mostly determined by the magnetocrystalline anisotropy parameters $k_1$ and $k_2$;
(iii) the slope of the resonance curve $f_{\rm r}(\mu_0 H)$ at $H\rightarrow0$ and the value $f_{\rm r}(H=0)$ are defined by the CPW-induced uniaxial anisotropy, i.e., by $k_{\rm a1}$ and $\alpha$ in Eq.~\ref{g_a1}.
Using the least-squares method for optimization through steps (i) to (iii) and back, after several runs, (iv) the fit is further optimized by adjusting parameter $k_{\rm u}$.
Following routines (i)-(iv) we have obtained an optimum fit of the G-line (see the red curve in Fig.~\ref{fit}a) with the following parameters: $4\pi M_{\rm s}=1876$~Oe, $k_1=-0.16$, $k_2=0.18$, $k_{\rm u}=-0.12$, $\theta=1.4$, $\phi_{\rm h}=126$, $k_{\rm a1}=0.025$, and $\alpha=177$.
Importantly, parameters of the cubic magnetocrystalline anisotropy $k_1$ and $k_2$ are by a factor of 2-3 higher than typical values at room temperature \cite{Lee_JAP_120_033905,Beaulieu_IEEEMagLett_9_3706005}.
This trend correlates well with the temperature dependence of the cubic magnetocrystalline anisotropy in YIG bulk single crystals \cite{Hansen_JAP_45_3638}.

After fitting the G-line, which corresponds to the FMR response of YIG areas at the CPW gaps, the only option to fit the C-line, which corresponds to the FMR response of YIG areas under the central CPW line, is to introduce an additional term into the energy $g$ that represents the second order uniaxial anisotropy induced by the CPW.
The term of the in-plane uniaxial anisotropy of the second order in the coordinates of Fig.~\ref{Coord} is 
\begin{eqnarray}
g_{\rm a}=-(k_{\rm a1}+2k_{\rm a2})\sin(\psi)^2\cos(\phi_{\rm h}-\alpha)^2+\nonumber
\\
+k_{\rm a2}\sin(\psi)^4\cos(\phi_{\rm h}-\alpha)^4
\label{g_a2}
\end{eqnarray}
Using magnetic parameters obtained for the G-line, the fitting procedure for the C-line with the anisotropy given by Eq.~\ref{g_a2} provides $k_{\rm a1}=0.121$ and $k_{\rm a2}=-0.048$.
This fit is shown in Fig.~\ref{fit}a with the blue curve.

Possible origins of the CPW-induced anisotropy include a distinct directionality of microwave currents.
Also, directionality of the surface stress can be considered that appears from differences in thermal expansion of narrow central transmission line of metal CPW and YIG/GGG oxides.
The surface stress may appear either due to deposition of Nb film at elevated temperature or due to performance of experiments at cryogenic temperatures.
For instance, the difference in thermal expansions between Nb and garnets can enable a strain in YIG at 2~K of up to $\epsilon\approx+6\times10^{-4}$ along the CPW in case of absence of mechanical relaxation in Nb.
In contrast, if a complete relaxation of tensions occurs in Nb at room temperature, the difference in thermal expansions enables the opposite-sign strain in YIG of $\epsilon\approx-4\times10^{-4}$.
Both values of the strain are well comparable with the growth induced tensions provided by the lattice misfit between the GGG substrate and YIG film that induces the uniaxial anisotropy in LPE-grown\cite{Dubs_JPDAP_50_204005} and PLD-grown films\cite{Howe_IEEEMagLet_6_3500504,Bhoi_JAP_123_203902}.
See the Appendix section for details.
Importantly, presence of both first- and second-order anisotropies suggests different mechanisms for their induction.

Figure~\ref{fit}b shows dependencies of orientations of magnetization $\psi(\mu_0 H)$ and $\phi_{\rm m}(\mu_0 H)$ for C- and G-FMR lines on magnetic field. 
A marginal difference between C- and G-curves in the entire field range  indicates co-alignment of magnetization orientations at both gap and center areas of the CPW, implying that the entire volume of YIG that is subjected to the FMR remains in the single-domain state throughout the experiment.

Our experimental setup does not allow us to study microwave transmission at higher temperatures $T\gtrsim15$~K.
Therefore, temperature dependence of magnetic parameters of YIG is not addressed in this report and can be found elsewhere \cite{Maier-Flaig_PRB_95_214423,Haidar_JAP_117_17D119,Beaulieu_IEEEMagLett_9_3706005}.

\subsection{FMR at $T<T_{\rm c}$. Impact of the superconducting critical state.}

\begin{figure}[!ht]
\begin{center}
\includegraphics[width=0.95\columnwidth]{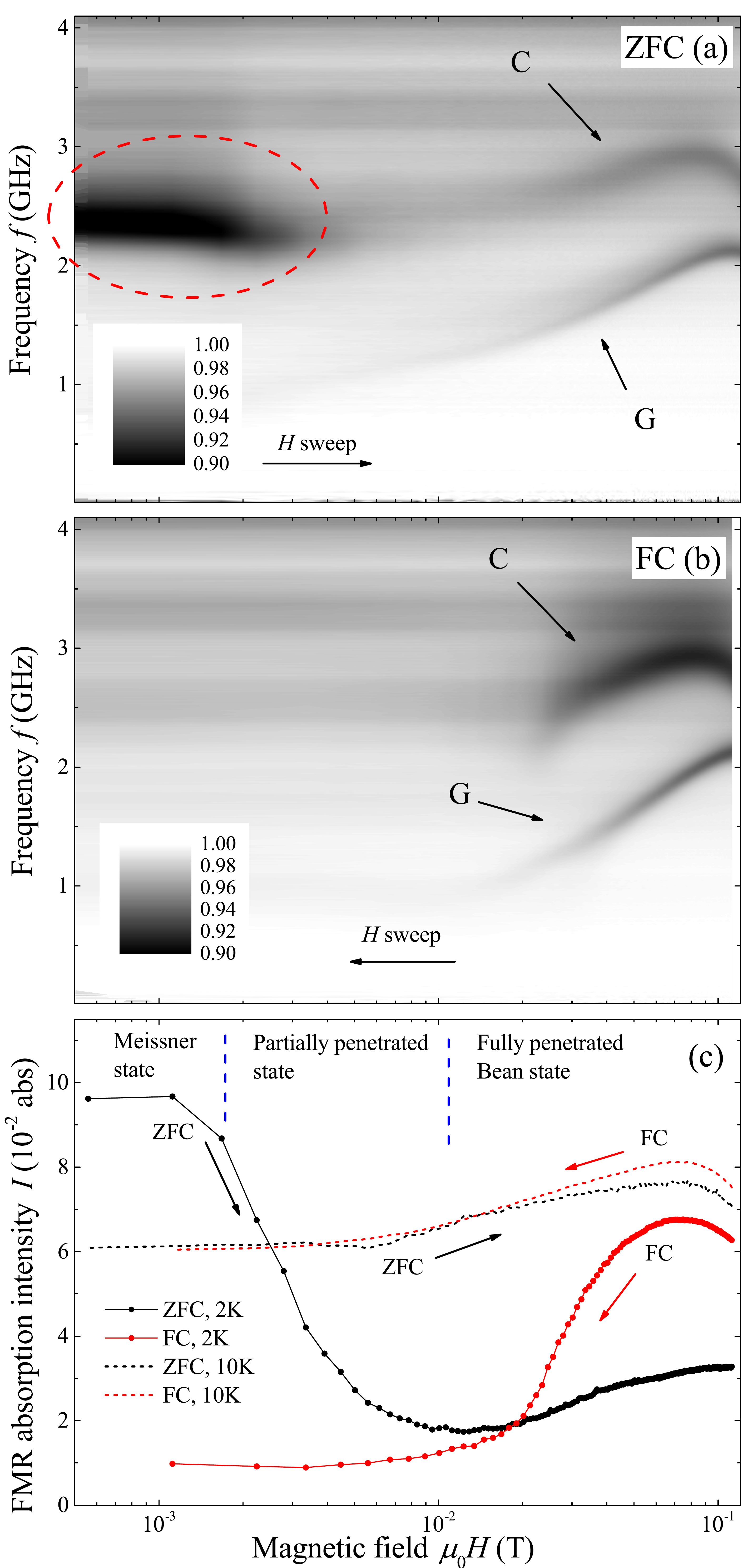}
\caption{Gray-scale-coded transmission spectra $|S_{21}(\mu_0 H,f)/S_{21}(\mu_0 H=0.5$~T$,f)|$ measured at 2~K starting from ZFC state (a) and FC state (b).
C and G spectral lines are indicated.
Red circle in (a) highlights FMR absorption in the Meissner state of the CPW.
c) Dependencies of resonance peak absorption for C-lines on magnetic field $I(\mu_0 H)$ obtained at 2~K and 10~K.
Direction of magnetic field sweep is indicated with arrows.
The $\mu_0 H$ axis is given on a log scale.
Field regions for three superconducting states of Nb at ZFC curve in (c) are separated with blue dashed lines.}
\label{hyst}
\end{center}
\end{figure}

At $T<T_{\rm c}$ of Nb, in presence of superconductivity, the FMR absorption spectrum changes (see Fig.~\ref{raw}b).
Since the Nb CPW is placed directly on top of the YIG film, all changes in absorptions in the C-branch can be attributed to the magnetization state under the Nb line. 
Therefore, the effect of superconductivity on the FMR can be tracked by analyzing the superconducting critical state of Nb film and its variation with applied magnetic field.  

Figure~\ref{hyst}a shows the zero-field-cooled (ZFC) transmission spectra that is acquired when the sample is cooled down to 2~K at zero magnetic field, and afterward $S_{21}$ measurements were performed while sweeping magnetic field from 0 to 0.11~T.
Figure~\ref{hyst}b shows the field-cooled (FC) transmission spectra that is acquired when the sample was cooled down to 2~K at $\mu_0 H=0.25$~T, and afterward $S_{21}$ measurements were performed while sweeping magnetic field back from 0.11~T down to 0.
The hysteresis in peak absorption can be tracked by fitting $S_{21}(f)$ curves at each value of $H$ and plotting dependencies of FMR amplitude $I$ on magnetic field $H$ (Fig.~\ref{hyst}c).
$I(\mu_0 H)$ dependency is caused by variation of the CPW-FMR coupling strength with magnetic field, i.e., by variation of magnetization and magnetic flux inhomogeneity in the YIG induced by the Nb superconducting critical state.
Note that no hysteresis in peak absorption is observed at $T>T_{\rm c}$ (Figs.~\ref{raw}a~and~\ref{hyst}c), where the transmission spectra is fully reversible and independent of the ZFC/FC initial state.

First we discuss the ZFC curve in Fig.~\ref{hyst}c, where three 
intervals in $I(\mu_0 H)$ can be distinguished.
At low fields, the strongest FMR absorption is observed with $I\sim0.1$ at $\mu_0 H$ up to $2\times10^{-3}$~T (highlighted with the red circle in Fig.~\ref{hyst}a).
This corresponds to the Meissner state of the Nb line when the Meissner screening currents circulate at the edges of the Nb film and exclude magnetic flux from its cross-section.
In the Meissner state, DC magnetic flux remains homogeneous across the Nb line and ensures a strong coupling of the CPW to the YIG at FMR.
At intermediate fields $2\times10^{-3}<\mu_0 H<10^{-2}$~T, the FMR absorption drops rapidly from $I\sim0.1$ to the minimum $I\sim0.02$,
caused by the partially penetrated superconducting critical state where superconducting vortices start to penetrate Nb film.
The magnetic flux profile in partially penetrated superconducting films is the most inhomogeneous \cite{Jooss_RPP_65_651,Wells_SUST_29_035014,Wells_SRep_7_40235}, which causes a weak coupling of the FMR to the CPW and low absorption intensity.
The partially penetrated state commences at the flux-focus enhanced first critical field of the superconducting film $\mu_0 H_{\rm c1}\sim2\times10^{-3}$~T, where the first Abrikosov vortices start to penetrate into the film, and terminates at the magnetic field of full penetration $10^{-2}$~T.
%
At high fields $\mu_0 H>0.01$~T, after full penetration is reached, magnetic flux in the superconducting film forms a constant gradient that can be depicted by the Bean critical state model \cite{BEAN_RMP_36_31,NORRIS_JPDAP_3_489,Chen_JAP_66_2489,Golovchanskiy_JAP_114_163910}.
The gradient is formed due to pinning of vortices and induces a homogeneous circulating critical currents.
Upon increasing magnetic field, both the pinning of vortices and the slope of magnetic flux reduce\cite{Chen_JAP_66_2489,Golovchanskiy_JAP_114_163910} making magnetic flux in YIG more homogenous.
A smaller gradient of the magnetic flux in the superconductor increases the coupling that we observe in gradual increase of the FMR peak absorption upon increasing magnetic field from 0.01~T to higher fields.
Note that such nonmonotonic behavior of $I(\mu_0 H)$ is not observed for the G-line (Fig.~\ref{hyst}a), which indicates additionally that the absorption at G-line is caused by the FMR at the gap areas of the CPW, where the influence of the superconducting state of Nb is marginal.

Increasing magnetic field further beyond the field range in Fig.~\ref{hyst}, the ZFC curve should coincide with the FC curve at the so-called irreversibility field\cite{Tinkham_PRL_61_1658,Golovchanskiy_SUST_29_075002,Golovchanskiy_JAP_114_163910}, where pinning of vortices becomes negligible.
The FC curve in Fig.~\ref{hyst} consists of two parts.
For $\mu_0 H>0.03$~T the coupling remains by a factor of $\sim2$ higher than one for the ZFC curve.
This difference is attributed to the fact that upon decreasing magnetic field, the Bean critical currents counter-act the Meissner currents, diamagnetic response of the superconducting film is reduced as compared to the ZFC measurement\cite{Golovchanskiy_SUST_29_075002},
and the influence on YIG at the FMR decreases.
Below 0.03~T, $I$ drops rapidly, which can be explained by a gradual formation of a complex remanent critical state at $H=0$ with highly nonuniformly distributed frozen magnetic flux.
Also at low magnetic fields, magnetization of individual Abrikosov vortices may contribute to YIG inhomogeneity by inducing substantial local magnetic fields of up to $\mu_0 H_{\rm v}\sim\Phi_0/\pi\lambda_{\rm L}^2\sim0.06$~T, where $\Phi_0$ is the magnetic flux quantum and $\lambda_{\rm L}\sim10^{-7}$~m is the typical London penetration depth in Nb films.
%

Overall, the influence of the superconducting critical state in our geometry on the FMR appears to be destructive.
The FMR intensity for both ZFC and FC curves remains below values of $I$ at $T>T_{\rm c}$ (Fig.~\ref{hyst}c).
However, magnetic hysteresis often is employed in magnetic logic devices.
Also, in vicinity to $H=0$, FMR is substantially stronger when superconductor is in the Meissner state than for normal metal CPW.
This effect may be a result of interaction of magnetic moments in YIG with Meissner screening currents in the ideal diamagnet.

\section{Conclusion}

In conclusion, ferromagnetic resonance of YIG film is studied in out-of-plane magnetic fields and cryogenic temperatures using a superconducting coplanar waveguide that is fabricated directly on top of the magnetic film (see Fig.~\ref{Sam}).
FMR absorption spectra are obtained in a wide field range.
Nonlinear dependence of the FMR frequency on magnetic field at low field values, below the field of saturation magnetization, showed a split of resonance into two spectral lines, which were identified as the FMR response of YIG at gap areas of the CPW and of YIG located directly under the central conducting line of the CPW.

A routine was developed for fitting the FMR lines.
This routine allowed us to obtain all magnetic parameters of YIG, i.e., the saturation magnetization, the gyromagnetic ratio, and parameters of magnetocrystalline and out-of-plane uniaxial anisotropies.
In addition, the fitting routine has issued the misalignment angle of 1.4$^{\circ}$ between magnetic field and the out-of-plane orientation, as well as parameters of in-plane magnetic anisotropy of the first and the second order, which are induced by the CPW.

The FMR spectrum at temperatures below the superconducting critical temperature of the waveguide showed a hysteresis in FMR peak absorption.
The hysteresis is explained by influence of magnetization of the Nb transmission line in the superconducting critical state.
Tracking the dependence of the intensity of the FMR on magnetic field allowed us to identify all fundamental states of a superconducting film in out-of-plane magnetic field, i.e., 
the Meissner state, the partially penetrated state, and the fully penetrated Bean critical state.
Also, it allowed explanation the hysteresis in the FMR absorption by the pinning of magnetic vortices, which induces the gradient of magnetic flux in superconducting films.
The gradient is controlled by direction of the magnetic field sweep. 

In general, this report suggests that development of magnonics at cryogenic temperatures may be beneficial due to:
(i)  substantially different  properties of magnetic materials, including magneto-crystalline anisotropy,
(ii) the possibility to engineer additional anisotropies with metal structures, and
(iii) the potential to affect the spectra by hybridization of a magnonic media with superconductors.
As a final remark we would like to point out a related work by Jeon et al. on the effect of the superconducting critical state on magnetization dynamics in thick superconductor/ferromagnet/superconductor trilayers \cite{Jeon_arXiv}.

\section{Acknowledgments}

The authors acknowledge Lucas Radtke and Yannick Schoen for assistance with sample preparation and initial measurements and Paul Baity for critical reading of the manuscript.
This work was supported by the European Research Council (ERC) under the Grant Agreement 648011, Deutsche Forschungsgemeinschaft (DFG) within Project INST 121384/138-1 FUGG.
C.D. thanks the Deutsche Forschungsgemeinschaft for financial support under contract number DFG DU 1427/2-1.
I.A.G. acknowledges support by the German Academic Exchange Service (DAAD) via the  program ``Research Stays for University Academics and Scientists 2017''.
I.A.G., N.N.A., V.V.R., and A.V.U. acknowledge the Ministry of Education and Science of the Russian Federation (Research Project K2-2018-015 in the framework of the Increase Competitiveness Program of NUST “MISiS”) for support in microwave measurements.
V.S.S., I.A.G., and D.S.B. acknowledge the Russian Science Foundation (RSF) (Project No. 18-72-10118) for support in numerical analysis and MFM investigations.
V.V.R. acknowledges a partial support by the Russian Foundation for Basic Research (RFBR) (Project No. 19-02-00316).

\section{Appendix: Stress induced in YIG by Nb CPW}

\begin{figure}[!ht]
\begin{center}
\includegraphics[width=0.95\columnwidth]{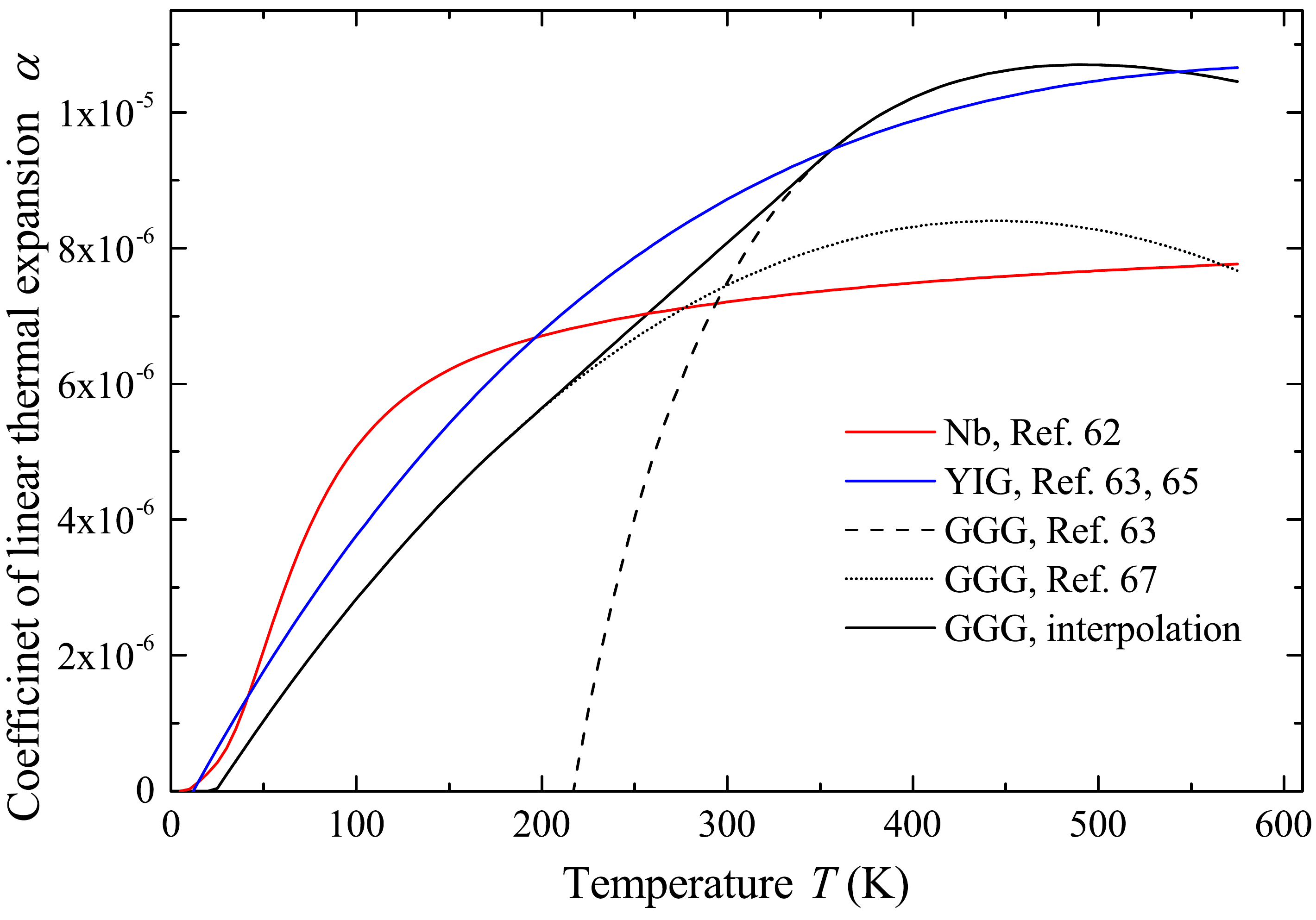}
\caption{
Dependencies of the thermal expansion coefficient on temperature $\alpha(T)$ for Nb, YIG and GGG.
}
\label{Exp}
\end{center}
\end{figure}

One possible cause of the CPW induced anisotropy that is derived in Sec.~IIIA is the stress in YIG that is forced by differences in thermal expansion of the narrow extended central transmission line of the metal CPW and YIG/GGG oxides.
Assuming that an unstressed continuous interface is formed between Nb and YIG during deposition of Nb at the deposition temperature $T_{\rm d}\approx600$~K,   
the stress at the interface at the measurement temperature $T_{\rm m}=2$~K can be estimated with the following expression 
\begin{equation}
\sigma\approx \frac{E}{1-\nu}\epsilon=\frac{E}{1-\nu}\int_{T_{\rm d}}^{T_{\rm m}}{[\alpha_{\rm G}(T)-\alpha_{\rm Nb}(T)]}dT,
\label{stress}
\end{equation}
where $\sigma$ is the stress in YIG, $E=2\times10^{12}$~dyne/cm$^2$ is the Young's modulus of YIG at the temperature range from 0 to 300~K\cite{GIBBONS_PR_110_770}, $\nu=0.29$ is the Poisson's ratio, $\epsilon$ is the strain at the interface at $T_{\rm m}$ due to the difference in thermal expansion, $\alpha_{\rm G}(T)$ and $\alpha_{\rm Nb}(T)$ are temperature dependencies of the linear thermal expansion of the garnet and Nb, respectively.
Importantly, the stress in Eq.~\ref{stress} implies absence of mechanical relaxation.

However, estimation of the stress at the Nb/YIG interface using Eq.~\ref{stress} is impeded. 
While thermo-mechanical properties of Nb are well studied in a wide temperature range\cite{Wang_MatSciEng_R23_101} from $\approx0$~K up to about the melting point, 
a consistent study of thermo-mechanical properties of YIG is not available for the required temperature range.
The coefficient $\alpha_{\rm G}(T)$ for YIG is available piecewise and can be obtained by interpolation of $\alpha_{\rm G}(T)$ at temperatures above\cite{GELLER_Jr_Appl_Cryst_2_86,Liang_Powder_Diffraction_14_2} and below\cite{Levinstein_JAP_37_2197} the room temperature.
On the other side, the coefficient $\alpha_{\rm G}(T)$ for YIG can be substituted with one for GGG since their thermo-mechanical properties are almost identical\cite{GELLER_Jr_Appl_Cryst_2_86,Liang_Powder_Diffraction_14_2}.
The coefficient $\alpha_{\rm G}(T)$ for GGG is reported for several temperature ranges separately: 
room temperature and higher temperature data is available in Refs.~\cite{GELLER_Jr_Appl_Cryst_2_86,Liang_Powder_Diffraction_14_2}, 
$\alpha_{\rm G}(T)$ at low temperatures is reported in Ref.~\cite{Antyukhov_Inorg_Mater_23_702} for the range from 6~K to 300~K and in Ref.~\cite{Fan_IEEE_13_448} for the range from 80~K to 330~K. 

Figure~\ref{Exp} shows dependencies of the thermal expansion on temperature $\alpha(T)$.
The red curve shows $\alpha_{\rm Nb}(T)$ for Nb that is calculated using Ref.~\cite{Wang_MatSciEng_R23_101}. 
The blue curve shows $\alpha_{\rm G}(T)$ for YIG that is calculated using Refs.~\cite{GELLER_Jr_Appl_Cryst_2_86,Levinstein_JAP_37_2197}.
Black dashed and dotted curves show $\alpha_{\rm G}(T)$ for GGG that are calculated using Ref.~\cite{GELLER_Jr_Appl_Cryst_2_86} and Ref.~\cite{Fan_IEEE_13_448}, respectively.
Solid black curve shows linear interpolation between lower-temperature and higher-temperature curves $\alpha_{\rm G}(T)$ for GGG at the range from 180~K up to 330~K.
The interpolated dependence for $\alpha_{\rm G}(T)$ is used for calculations.

Calculations with Eq.~\ref{stress} and coefficients $\alpha(T)$ in Fig.~\ref{Exp} provide the strain at YIG/Nb interface  $\epsilon\approx+6.4\times10^{-4}$ that produces a compressive stress $\sigma\sim10^9$~dyne/cm$^2$.
Note however, that if the room-temperature deposition of Nb takes place, or the strain in Nb relaxes at room temperature, according to Eq.~\ref{stress} and Fig.~\ref{Exp} an opposite-sign strain $\epsilon\approx-4\times10^{-5}$ emerges at cryogenic temperature $T_{\rm m}$.
If the data for GGG is used instead of YIG, the integral in Eq.~\ref{stress} provides approximately the same strain $\epsilon\approx+5.6\times10^{-4}$ at the interface with unrelaxed Nb, and a larger opposite-sign strain $\epsilon\approx-4\times10^{-4}$ at the interface with the room-$T$ deposited or relaxed Nb.
These values are well comparable with the growth induced tensions provided by the lattice misfit between the GGG substrate and the YIG film that induces the uniaxial anisotropy in LPE-grown\cite{Dubs_JPDAP_50_204005} and PLD-grown\cite{Howe_IEEEMagLet_6_3500504,Bhoi_JAP_123_203902} films.

\bibliography{A_MyBib_YIGpaper_titles}

\begin{thebibliography}{68}%
\makeatletter
\providecommand \@ifxundefined [1]{%
 \@ifx{#1\undefined}
}%
\providecommand \@ifnum [1]{%
 \ifnum #1\expandafter \@firstoftwo
 \else \expandafter \@secondoftwo
 \fi
}%
\providecommand \@ifx [1]{%
 \ifx #1\expandafter \@firstoftwo
 \else \expandafter \@secondoftwo
 \fi
}%
\providecommand \natexlab [1]{#1}%
\providecommand \enquote  [1]{``#1''}%
\providecommand \bibnamefont  [1]{#1}%
\providecommand \bibfnamefont [1]{#1}%
\providecommand \citenamefont [1]{#1}%
\providecommand \href@noop [0]{\@secondoftwo}%
\providecommand \href [0]{\begingroup \@sanitize@url \@href}%
\providecommand \@href[1]{\@@startlink{#1}\@@href}%
\providecommand \@@href[1]{\endgroup#1\@@endlink}%
\providecommand \@sanitize@url [0]{\catcode `\\12\catcode `\$12\catcode
  `\&12\catcode `\#12\catcode `\^12\catcode `\_12\catcode `\%12\relax}%
\providecommand \@@startlink[1]{}%
\providecommand \@@endlink[0]{}%
\providecommand \url  [0]{\begingroup\@sanitize@url \@url }%
\providecommand \@url [1]{\endgroup\@href {#1}{\urlprefix }}%
\providecommand \urlprefix  [0]{URL }%
\providecommand \Eprint [0]{\href }%
\providecommand \doibase [0]{http://dx.doi.org/}%
\providecommand \selectlanguage [0]{\@gobble}%
\providecommand \bibinfo  [0]{\@secondoftwo}%
\providecommand \bibfield  [0]{\@secondoftwo}%
\providecommand \translation [1]{[#1]}%
\providecommand \BibitemOpen [0]{}%
\providecommand \bibitemStop [0]{}%
\providecommand \bibitemNoStop [0]{.\EOS\space}%
\providecommand \EOS [0]{\spacefactor3000\relax}%
\providecommand \BibitemShut  [1]{\csname bibitem#1\endcsname}%
\let\auto@bib@innerbib\@empty
\bibitem [{\citenamefont {Lenk}\ \emph {et~al.}(2011)\citenamefont {Lenk},
  \citenamefont {Ulrichs}, \citenamefont {Garbs},\ and\ \citenamefont
  {M{\"u}nzenberg}}]{Lenk_PR_507_107}%
  \BibitemOpen
  \bibfield  {author} {\bibinfo {author} {\bibfnamefont {B.}~\bibnamefont
  {Lenk}}, \bibinfo {author} {\bibfnamefont {H.}~\bibnamefont {Ulrichs}},
  \bibinfo {author} {\bibfnamefont {F.}~\bibnamefont {Garbs}}, \ and\ \bibinfo
  {author} {\bibfnamefont {M.}~\bibnamefont {M{\"u}nzenberg}},\ }\bibfield
  {title} {\enquote {\bibinfo {title} {The building blocks of magnonics},}\
  }\href@noop {} {\bibfield  {journal} {\bibinfo  {journal} {Phys. Rep.}\
  }\textbf {\bibinfo {volume} {507}},\ \bibinfo {pages} {107} (\bibinfo {year}
  {2011})}\BibitemShut {NoStop}%
\bibitem [{\citenamefont {Chumak}\ \emph {et~al.}(2015)\citenamefont {Chumak},
  \citenamefont {Vasyuchka}, \citenamefont {Serga},\ and\ \citenamefont
  {Hillebrands}}]{Chumak_NatPhys_11_453}%
  \BibitemOpen
  \bibfield  {author} {\bibinfo {author} {\bibfnamefont {A.~V.}\ \bibnamefont
  {Chumak}}, \bibinfo {author} {\bibfnamefont {V.~I.}\ \bibnamefont
  {Vasyuchka}}, \bibinfo {author} {\bibfnamefont {A.~A.}\ \bibnamefont
  {Serga}}, \ and\ \bibinfo {author} {\bibfnamefont {B.}~\bibnamefont
  {Hillebrands}},\ }\bibfield  {title} {\enquote {\bibinfo {title} {Magnon
  spintronics},}\ }\href@noop {} {\bibfield  {journal} {\bibinfo  {journal}
  {Nat. Phys.}\ }\textbf {\bibinfo {volume} {11}},\ \bibinfo {pages} {453}
  (\bibinfo {year} {2015})}\BibitemShut {NoStop}%
\bibitem [{\citenamefont {Kajiwara}\ \emph {et~al.}(2010)\citenamefont
  {Kajiwara}, \citenamefont {Harii}, \citenamefont {Takahashi}, \citenamefont
  {Ohe}, \citenamefont {Uchida}, \citenamefont {Mizuguchi}, \citenamefont
  {Umezawa}, \citenamefont {Kawai}, \citenamefont {Ando}, \citenamefont
  {Takanashi}, \citenamefont {Maekawa},\ and\ \citenamefont
  {Saitoh}}]{Kajiwara_Nat_464_262}%
  \BibitemOpen
  \bibfield  {author} {\bibinfo {author} {\bibfnamefont {Y.}~\bibnamefont
  {Kajiwara}}, \bibinfo {author} {\bibfnamefont {K.}~\bibnamefont {Harii}},
  \bibinfo {author} {\bibfnamefont {S.}~\bibnamefont {Takahashi}}, \bibinfo
  {author} {\bibfnamefont {J.}~\bibnamefont {Ohe}}, \bibinfo {author}
  {\bibfnamefont {K.}~\bibnamefont {Uchida}}, \bibinfo {author} {\bibfnamefont
  {M.}~\bibnamefont {Mizuguchi}}, \bibinfo {author} {\bibfnamefont
  {H.}~\bibnamefont {Umezawa}}, \bibinfo {author} {\bibfnamefont
  {H.}~\bibnamefont {Kawai}}, \bibinfo {author} {\bibfnamefont
  {K.}~\bibnamefont {Ando}}, \bibinfo {author} {\bibfnamefont {K.}~\bibnamefont
  {Takanashi}}, \bibinfo {author} {\bibfnamefont {S.}~\bibnamefont {Maekawa}},
  \ and\ \bibinfo {author} {\bibfnamefont {E.}~\bibnamefont {Saitoh}},\
  }\bibfield  {title} {\enquote {\bibinfo {title} {Transmission of electrical
  signals by spin-wave interconversion in a magnetic insulator},}\ }\href@noop
  {} {\bibfield  {journal} {\bibinfo  {journal} {Nature}\ }\textbf {\bibinfo
  {volume} {464}},\ \bibinfo {pages} {262} (\bibinfo {year}
  {2010})}\BibitemShut {NoStop}%
\bibitem [{Spe(2017)}]{Spec_Iss}%
  \BibitemOpen
  \href@noop {} {\bibfield  {journal} {\bibinfo  {journal} {Special Issue on
  Magnonics, J. Phys. D: Appl. Phys.}\ }\textbf {\bibinfo {volume} {50}}
  (\bibinfo {year} {2017})}\BibitemShut {NoStop}%
\bibitem [{\citenamefont {Demokritov}\ and\ \citenamefont
  {Slavin}(2013)}]{Demokritov_book}%
  \BibitemOpen
  \bibinfo {editor} {\bibfnamefont {S.~O.}\ \bibnamefont {Demokritov}}\ and\
  \bibinfo {editor} {\bibfnamefont {A.~N.}\ \bibnamefont {Slavin}},\ eds.,\
  \href@noop {} {\emph {\bibinfo {title} {Magnonics: From Fundamentals to
  Applications}}}\ (\bibinfo  {publisher} {Springer-Verlag Berlin Heidelberg},\
  \bibinfo {year} {2013})\BibitemShut {NoStop}%
\bibitem [{\citenamefont {Evelt}\ \emph {et~al.}(2018)\citenamefont {Evelt},
  \citenamefont {Soumah}, \citenamefont {Rinkevich}, \citenamefont
  {Demokritov}, \citenamefont {Anane}, \citenamefont {Cros}, \citenamefont
  {Youssef}, \citenamefont {de~Loubens}, \citenamefont {Klein}, \citenamefont
  {Bortolotti},\ and\ \citenamefont {Demidov}}]{Evelt_PRAppl_10_041002}%
  \BibitemOpen
  \bibfield  {author} {\bibinfo {author} {\bibfnamefont {M.}~\bibnamefont
  {Evelt}}, \bibinfo {author} {\bibfnamefont {L.}~\bibnamefont {Soumah}},
  \bibinfo {author} {\bibfnamefont {A.B.}\ \bibnamefont {Rinkevich}}, \bibinfo
  {author} {\bibfnamefont {S.O.}\ \bibnamefont {Demokritov}}, \bibinfo {author}
  {\bibfnamefont {A.}~\bibnamefont {Anane}}, \bibinfo {author} {\bibfnamefont
  {V.}~\bibnamefont {Cros}}, \bibinfo {author} {\bibfnamefont {Jamal~Ben}\
  \bibnamefont {Youssef}}, \bibinfo {author} {\bibfnamefont {G.}~\bibnamefont
  {de~Loubens}}, \bibinfo {author} {\bibfnamefont {O.}~\bibnamefont {Klein}},
  \bibinfo {author} {\bibfnamefont {P.}~\bibnamefont {Bortolotti}}, \ and\
  \bibinfo {author} {\bibfnamefont {V.E.}\ \bibnamefont {Demidov}},\ }\bibfield
   {title} {\enquote {\bibinfo {title} {Emission of coherent propagating
  magnons by insulator-based spin-orbit-torque oscillators},}\ }\href@noop {}
  {\bibfield  {journal} {\bibinfo  {journal} {Phys. Rev. Appl.}\ }\textbf
  {\bibinfo {volume} {10}},\ \bibinfo {pages} {041002} (\bibinfo {year}
  {2018})}\BibitemShut {NoStop}%
\bibitem [{\citenamefont {Huebl}\ \emph {et~al.}(2013)\citenamefont {Huebl},
  \citenamefont {Zollitsch}, \citenamefont {Lotze}, \citenamefont {Hocke},
  \citenamefont {Greifenstein}, \citenamefont {Marx}, \citenamefont {Gross},\
  and\ \citenamefont {Goennenwein}}]{Huebl_PRL_111_127003}%
  \BibitemOpen
  \bibfield  {author} {\bibinfo {author} {\bibfnamefont {H.}~\bibnamefont
  {Huebl}}, \bibinfo {author} {\bibfnamefont {C.~W.}\ \bibnamefont
  {Zollitsch}}, \bibinfo {author} {\bibfnamefont {J.}~\bibnamefont {Lotze}},
  \bibinfo {author} {\bibfnamefont {F.}~\bibnamefont {Hocke}}, \bibinfo
  {author} {\bibfnamefont {M.}~\bibnamefont {Greifenstein}}, \bibinfo {author}
  {\bibfnamefont {A.}~\bibnamefont {Marx}}, \bibinfo {author} {\bibfnamefont
  {R.}~\bibnamefont {Gross}}, \ and\ \bibinfo {author} {\bibfnamefont
  {S.~T.~B.}\ \bibnamefont {Goennenwein}},\ }\bibfield  {title} {\enquote
  {\bibinfo {title} {High cooperativity in coupled microwave resonator
  ferrimagnetic insulator hybrids},}\ }\href@noop {} {\bibfield  {journal}
  {\bibinfo  {journal} {Phys. Rev. Lett.}\ }\textbf {\bibinfo {volume} {111}},\
  \bibinfo {pages} {127003} (\bibinfo {year} {2013})}\BibitemShut {NoStop}%
\bibitem [{\citenamefont {Tabuchi}\ \emph {et~al.}(2014)\citenamefont
  {Tabuchi}, \citenamefont {Ishino}, \citenamefont {Ishikawa}, \citenamefont
  {Yamazaki}, \citenamefont {Usami},\ and\ \citenamefont
  {Nakamura}}]{Tabuchi_PRL_113_083603}%
  \BibitemOpen
  \bibfield  {author} {\bibinfo {author} {\bibfnamefont {Y.}~\bibnamefont
  {Tabuchi}}, \bibinfo {author} {\bibfnamefont {S.}~\bibnamefont {Ishino}},
  \bibinfo {author} {\bibfnamefont {T.}~\bibnamefont {Ishikawa}}, \bibinfo
  {author} {\bibfnamefont {R.}~\bibnamefont {Yamazaki}}, \bibinfo {author}
  {\bibfnamefont {K.}~\bibnamefont {Usami}}, \ and\ \bibinfo {author}
  {\bibfnamefont {Y.}~\bibnamefont {Nakamura}},\ }\bibfield  {title} {\enquote
  {\bibinfo {title} {Hybridizing ferromagnetic magnons and microwave photons in
  the quantum limit},}\ }\href@noop {} {\bibfield  {journal} {\bibinfo
  {journal} {Phys. Rev. Lett.}\ }\textbf {\bibinfo {volume} {113}},\ \bibinfo
  {pages} {083603} (\bibinfo {year} {2014})}\BibitemShut {NoStop}%
\bibitem [{\citenamefont {Zhang}\ \emph {et~al.}(2014)\citenamefont {Zhang},
  \citenamefont {Zou}, \citenamefont {Jiang},\ and\ \citenamefont
  {Tang}}]{Zhang_PRL_113_156401}%
  \BibitemOpen
  \bibfield  {author} {\bibinfo {author} {\bibfnamefont {X.}~\bibnamefont
  {Zhang}}, \bibinfo {author} {\bibfnamefont {C.-L.}\ \bibnamefont {Zou}},
  \bibinfo {author} {\bibfnamefont {L.}~\bibnamefont {Jiang}}, \ and\ \bibinfo
  {author} {\bibfnamefont {H.~X.}\ \bibnamefont {Tang}},\ }\bibfield  {title}
  {\enquote {\bibinfo {title} {Strongly coupled magnons and cavity microwave
  photons},}\ }\href@noop {} {\bibfield  {journal} {\bibinfo  {journal} {Phys.
  Rev. Lett.}\ }\textbf {\bibinfo {volume} {113}},\ \bibinfo {pages} {156401}
  (\bibinfo {year} {2014})}\BibitemShut {NoStop}%
\bibitem [{\citenamefont {Morris}\ \emph {et~al.}(2017)\citenamefont {Morris},
  \citenamefont {van Loo}, \citenamefont {Kosen},\ and\ \citenamefont
  {Karenowska}}]{Morris_SciRep_7_11511}%
  \BibitemOpen
  \bibfield  {author} {\bibinfo {author} {\bibfnamefont {R.~G.~E.}\
  \bibnamefont {Morris}}, \bibinfo {author} {\bibfnamefont {A.~F.}\
  \bibnamefont {van Loo}}, \bibinfo {author} {\bibfnamefont {S.}~\bibnamefont
  {Kosen}}, \ and\ \bibinfo {author} {\bibfnamefont {A.~D.}\ \bibnamefont
  {Karenowska}},\ }\bibfield  {title} {\enquote {\bibinfo {title} {Strong
  coupling of magnons in a {YIG} sphere to photons in a planar superconducting
  resonator in the quantum limit},}\ }\href@noop {} {\bibfield  {journal}
  {\bibinfo  {journal} {Sci. Rep.}\ }\textbf {\bibinfo {volume} {7}},\ \bibinfo
  {pages} {11511} (\bibinfo {year} {2017})}\BibitemShut {NoStop}%
\bibitem [{\citenamefont {Pfirrmann}\ \emph {et~al.}(2019)\citenamefont
  {Pfirrmann}, \citenamefont {Boventer}, \citenamefont {Schneider},
  \citenamefont {Kl{\"a}ui}, \citenamefont {Ustinov},\ and\ \citenamefont
  {Weides}}]{Weides}%
  \BibitemOpen
  \bibfield  {author} {\bibinfo {author} {\bibfnamefont {Marco}\ \bibnamefont
  {Pfirrmann}}, \bibinfo {author} {\bibfnamefont {Isabella}\ \bibnamefont
  {Boventer}}, \bibinfo {author} {\bibfnamefont {Andre}\ \bibnamefont
  {Schneider}}, \bibinfo {author} {\bibfnamefont {Tim Wolzand~Mathias}\
  \bibnamefont {Kl{\"a}ui}}, \bibinfo {author} {\bibfnamefont {Alexey~V.}\
  \bibnamefont {Ustinov}}, \ and\ \bibinfo {author} {\bibfnamefont {Martin}\
  \bibnamefont {Weides}},\ }\href@noop {} {\enquote {\bibinfo {title} {Magnons
  at low excitations: Observation of incoherent coupling to a bath of
  two-level-systems},}\ } (\bibinfo {year} {2019}),\ \Eprint
  {http://arxiv.org/abs/arXiv:1903.03981} {arXiv:1903.03981} \BibitemShut
  {NoStop}%
\bibitem [{\citenamefont {Golovchanskiy}\ \emph
  {et~al.}(2018{\natexlab{a}})\citenamefont {Golovchanskiy}, \citenamefont
  {Abramov}, \citenamefont {Stolyarov}, \citenamefont {Shchetinin},
  \citenamefont {Dzhumaev}, \citenamefont {Averkin}, \citenamefont {Kozlov},
  \citenamefont {Golubov}, \citenamefont {Ryazanov},\ and\ \citenamefont
  {Ustinov}}]{Golovchanskiy_JAP_123_173904}%
  \BibitemOpen
  \bibfield  {author} {\bibinfo {author} {\bibfnamefont {I.~A.}\ \bibnamefont
  {Golovchanskiy}}, \bibinfo {author} {\bibfnamefont {N.~N.}\ \bibnamefont
  {Abramov}}, \bibinfo {author} {\bibfnamefont {V.~S.}\ \bibnamefont
  {Stolyarov}}, \bibinfo {author} {\bibfnamefont {I.~V.}\ \bibnamefont
  {Shchetinin}}, \bibinfo {author} {\bibfnamefont {P.~S.}\ \bibnamefont
  {Dzhumaev}}, \bibinfo {author} {\bibfnamefont {A.~S.}\ \bibnamefont
  {Averkin}}, \bibinfo {author} {\bibfnamefont {S.~N.}\ \bibnamefont {Kozlov}},
  \bibinfo {author} {\bibfnamefont {A.~A.}\ \bibnamefont {Golubov}}, \bibinfo
  {author} {\bibfnamefont {V.~V.}\ \bibnamefont {Ryazanov}}, \ and\ \bibinfo
  {author} {\bibfnamefont {A.~V.}\ \bibnamefont {Ustinov}},\ }\bibfield
  {title} {\enquote {\bibinfo {title} {Probing dynamics of micro-magnets with
  multi-mode superconducting resonator},}\ }\href@noop {} {\bibfield  {journal}
  {\bibinfo  {journal} {J. Appl. Phys.}\ }\textbf {\bibinfo {volume} {123}},\
  \bibinfo {pages} {173904} (\bibinfo {year} {2018}{\natexlab{a}})}\BibitemShut
  {NoStop}%
\bibitem [{\citenamefont {Barnes}\ \emph {et~al.}(2011)\citenamefont {Barnes},
  \citenamefont {Aprili}, \citenamefont {Petkovic},\ and\ \citenamefont
  {Maekawa}}]{Barnes_SUST_24_024020}%
  \BibitemOpen
  \bibfield  {author} {\bibinfo {author} {\bibfnamefont {S.~E.}\ \bibnamefont
  {Barnes}}, \bibinfo {author} {\bibfnamefont {M.}~\bibnamefont {Aprili}},
  \bibinfo {author} {\bibfnamefont {I.}~\bibnamefont {Petkovic}}, \ and\
  \bibinfo {author} {\bibfnamefont {S.}~\bibnamefont {Maekawa}},\ }\bibfield
  {title} {\enquote {\bibinfo {title} {Ferromagnetic resonance with a magnetic
  josephson junction},}\ }\href@noop {} {\bibfield  {journal} {\bibinfo
  {journal} {Supercond. Sci. Technol.}\ }\textbf {\bibinfo {volume} {24}},\
  \bibinfo {pages} {024020} (\bibinfo {year} {2011})}\BibitemShut {NoStop}%
\bibitem [{\citenamefont {Mai}\ \emph {et~al.}(2011)\citenamefont {Mai},
  \citenamefont {Kandelaki}, \citenamefont {Volkov},\ and\ \citenamefont
  {Efetov}}]{Mai_PRB_84_144519}%
  \BibitemOpen
  \bibfield  {author} {\bibinfo {author} {\bibfnamefont {S.}~\bibnamefont
  {Mai}}, \bibinfo {author} {\bibfnamefont {E.}~\bibnamefont {Kandelaki}},
  \bibinfo {author} {\bibfnamefont {A.~F.}\ \bibnamefont {Volkov}}, \ and\
  \bibinfo {author} {\bibfnamefont {K.~B.}\ \bibnamefont {Efetov}},\ }\bibfield
   {title} {\enquote {\bibinfo {title} {Interaction of josephson and magnetic
  oscillations in josephson tunnel junctions with a ferromagnetic layer},}\
  }\href@noop {} {\bibfield  {journal} {\bibinfo  {journal} {Phys. Rev. B}\
  }\textbf {\bibinfo {volume} {84}},\ \bibinfo {pages} {144519} (\bibinfo
  {year} {2011})}\BibitemShut {NoStop}%
\bibitem [{\citenamefont {Golovchanskiy}\ \emph {et~al.}(2017)\citenamefont
  {Golovchanskiy}, \citenamefont {Abramov}, \citenamefont {Stolyarov},
  \citenamefont {Emelyanova}, \citenamefont {Golubov}, \citenamefont
  {Ustinov},\ and\ \citenamefont {Ryazanov}}]{Golovchanskiy_SUST_30_054005}%
  \BibitemOpen
  \bibfield  {author} {\bibinfo {author} {\bibfnamefont {I.~A.}\ \bibnamefont
  {Golovchanskiy}}, \bibinfo {author} {\bibfnamefont {N.~N.}\ \bibnamefont
  {Abramov}}, \bibinfo {author} {\bibfnamefont {V.~S.}\ \bibnamefont
  {Stolyarov}}, \bibinfo {author} {\bibfnamefont {O.~V.}\ \bibnamefont
  {Emelyanova}}, \bibinfo {author} {\bibfnamefont {A.~A.}\ \bibnamefont
  {Golubov}}, \bibinfo {author} {\bibfnamefont {A.~V.}\ \bibnamefont
  {Ustinov}}, \ and\ \bibinfo {author} {\bibfnamefont {V.~V.}\ \bibnamefont
  {Ryazanov}},\ }\bibfield  {title} {\enquote {\bibinfo {title} {Ferromagnetic
  resonance with long josephson junction},}\ }\href@noop {} {\bibfield
  {journal} {\bibinfo  {journal} {Supercond. Sci. Technol.}\ }\textbf {\bibinfo
  {volume} {30}},\ \bibinfo {pages} {054005} (\bibinfo {year}
  {2017})}\BibitemShut {NoStop}%
\bibitem [{\citenamefont {Lebed}\ and\ \citenamefont
  {Yzkovlev}(1989)}]{Lebed_PisZTF_15_27}%
  \BibitemOpen
  \bibfield  {author} {\bibinfo {author} {\bibfnamefont {B.M.}\ \bibnamefont
  {Lebed}}\ and\ \bibinfo {author} {\bibfnamefont {S.V.}\ \bibnamefont
  {Yzkovlev}},\ }\bibfield  {title} {\enquote {\bibinfo {title} {Dispersion of
  surface spin waves in a layered superconductor-ferrite structure},}\
  }\href@noop {} {\bibfield  {journal} {\bibinfo  {journal} {Pis'ma v ZhTF (in
  Russian)}\ }\textbf {\bibinfo {volume} {15(14)}},\ \bibinfo {pages} {27}
  (\bibinfo {year} {1989})}\BibitemShut {NoStop}%
\bibitem [{\citenamefont {Anfinogenov}\ \emph {et~al.}(1989)\citenamefont
  {Anfinogenov}, \citenamefont {Gulyaev}, \citenamefont {Zilberman},
  \citenamefont {Kotelyanskiy}, \citenamefont {Polzikova},\ and\ \citenamefont
  {Suhanov}}]{Anfinogenov_PisZTF_15_24}%
  \BibitemOpen
  \bibfield  {author} {\bibinfo {author} {\bibfnamefont {V.~B.}\ \bibnamefont
  {Anfinogenov}}, \bibinfo {author} {\bibfnamefont {Y.~V.}\ \bibnamefont
  {Gulyaev}}, \bibinfo {author} {\bibfnamefont {P.~E.}\ \bibnamefont
  {Zilberman}}, \bibinfo {author} {\bibfnamefont {I.~M.}\ \bibnamefont
  {Kotelyanskiy}}, \bibinfo {author} {\bibfnamefont {N.~I.}\ \bibnamefont
  {Polzikova}}, \ and\ \bibinfo {author} {\bibfnamefont {A.~A.}\ \bibnamefont
  {Suhanov}},\ }\bibfield  {title} {\enquote {\bibinfo {title} {Observation of
  the electronic absorption of magnetostatic waves in a ferrite - high
  temperature superconductor structure},}\ }\href@noop {} {\bibfield  {journal}
  {\bibinfo  {journal} {Pis'ma v ZhTF (in Russian)}\ }\textbf {\bibinfo
  {volume} {15(19)}},\ \bibinfo {pages} {24} (\bibinfo {year}
  {1989})}\BibitemShut {NoStop}%
\bibitem [{\citenamefont {Golovchanskiy}\ \emph
  {et~al.}(2018{\natexlab{b}})\citenamefont {Golovchanskiy}, \citenamefont
  {Abramov}, \citenamefont {Stolyarov}, \citenamefont {Bolginov}, \citenamefont
  {Ryazanov}, \citenamefont {Golubov},\ and\ \citenamefont
  {Ustinov}}]{Golovchanskiy_AFM_28_1802375}%
  \BibitemOpen
  \bibfield  {author} {\bibinfo {author} {\bibfnamefont {I.~A.}\ \bibnamefont
  {Golovchanskiy}}, \bibinfo {author} {\bibfnamefont {N.~N.}\ \bibnamefont
  {Abramov}}, \bibinfo {author} {\bibfnamefont {V.~S.}\ \bibnamefont
  {Stolyarov}}, \bibinfo {author} {\bibfnamefont {V.~V.}\ \bibnamefont
  {Bolginov}}, \bibinfo {author} {\bibfnamefont {V.~V.}\ \bibnamefont
  {Ryazanov}}, \bibinfo {author} {\bibfnamefont {A.~A.}\ \bibnamefont
  {Golubov}}, \ and\ \bibinfo {author} {\bibfnamefont {A.~V.}\ \bibnamefont
  {Ustinov}},\ }\bibfield  {title} {\enquote {\bibinfo {title}
  {Ferromagnet/superconductor hybridization for magnonic applications},}\
  }\href@noop {} {\bibfield  {journal} {\bibinfo  {journal} {Adv. Func.
  Mater.}\ }\textbf {\bibinfo {volume} {28}},\ \bibinfo {pages} {1802375}
  (\bibinfo {year} {2018}{\natexlab{b}})}\BibitemShut {NoStop}%
\bibitem [{\citenamefont {Golovchanskiy}\ \emph
  {et~al.}(2018{\natexlab{c}})\citenamefont {Golovchanskiy}, \citenamefont
  {Abramov}, \citenamefont {Stolyarov}, \citenamefont {Ryazanov}, \citenamefont
  {Golubov},\ and\ \citenamefont {Ustinov}}]{Golovchanskiy_PRB_sub}%
  \BibitemOpen
  \bibfield  {author} {\bibinfo {author} {\bibfnamefont {I.~A.}\ \bibnamefont
  {Golovchanskiy}}, \bibinfo {author} {\bibfnamefont {N.~N.}\ \bibnamefont
  {Abramov}}, \bibinfo {author} {\bibfnamefont {V.~S.}\ \bibnamefont
  {Stolyarov}}, \bibinfo {author} {\bibfnamefont {V.~V.}\ \bibnamefont
  {Ryazanov}}, \bibinfo {author} {\bibfnamefont {A.~A.}\ \bibnamefont
  {Golubov}}, \ and\ \bibinfo {author} {\bibfnamefont {A.~V.}\ \bibnamefont
  {Ustinov}},\ }\bibfield  {title} {\enquote {\bibinfo {title} {Modified
  dispersion law for spin waves coupled to a superconductor},}\ }\href@noop {}
  {\bibfield  {journal} {\bibinfo  {journal} {J. Appl. Phys.}\ }\textbf
  {\bibinfo {volume} {124}},\ \bibinfo {pages} {233903} (\bibinfo {year}
  {2018}{\natexlab{c}})}\BibitemShut {NoStop}%
\bibitem [{\citenamefont {Dobrovolskiy}\ \emph {et~al.}(2019)\citenamefont
  {Dobrovolskiy}, \citenamefont {Sachser}, \citenamefont {Br{\"a}cher},
  \citenamefont {B{\"o}ttcher}, \citenamefont {Kruglyak}, \citenamefont {Vovk},
  \citenamefont {Shklovskij}, \citenamefont {Huth}, \citenamefont
  {Hillebrands},\ and\ \citenamefont {Chumak}}]{Dobrovolskiy_NatPhys}%
  \BibitemOpen
  \bibfield  {author} {\bibinfo {author} {\bibfnamefont {O.~V.}\ \bibnamefont
  {Dobrovolskiy}}, \bibinfo {author} {\bibfnamefont {R.}~\bibnamefont
  {Sachser}}, \bibinfo {author} {\bibfnamefont {T.}~\bibnamefont
  {Br{\"a}cher}}, \bibinfo {author} {\bibfnamefont {T.}~\bibnamefont
  {B{\"o}ttcher}}, \bibinfo {author} {\bibfnamefont {V.~V.}\ \bibnamefont
  {Kruglyak}}, \bibinfo {author} {\bibfnamefont {R.~V.}\ \bibnamefont {Vovk}},
  \bibinfo {author} {\bibfnamefont {V.~A.}\ \bibnamefont {Shklovskij}},
  \bibinfo {author} {\bibfnamefont {M.}~\bibnamefont {Huth}}, \bibinfo {author}
  {\bibfnamefont {B.}~\bibnamefont {Hillebrands}}, \ and\ \bibinfo {author}
  {\bibfnamefont {A.~V.}\ \bibnamefont {Chumak}},\ }\bibfield  {title}
  {\enquote {\bibinfo {title} {Magnon-fluxon interaction in a
  ferromagnet/superconductor heterostructure},}\ }\href@noop {} {\bibfield
  {journal} {\bibinfo  {journal} {Nature Physics}\ ,\ \bibinfo {pages}
  {doi.org/10.1038/s41567--019--0428--5}} (\bibinfo {year} {2019})}\BibitemShut
  {NoStop}%
\bibitem [{\citenamefont {Pimenov}\ \emph {et~al.}(2005)\citenamefont
  {Pimenov}, \citenamefont {Loidl}, \citenamefont {Przyslupski},\ and\
  \citenamefont {Dabrowski}}]{Pimenov_PRL_95_247009}%
  \BibitemOpen
  \bibfield  {author} {\bibinfo {author} {\bibfnamefont {A.}~\bibnamefont
  {Pimenov}}, \bibinfo {author} {\bibfnamefont {A.}~\bibnamefont {Loidl}},
  \bibinfo {author} {\bibfnamefont {P.}~\bibnamefont {Przyslupski}}, \ and\
  \bibinfo {author} {\bibfnamefont {B.}~\bibnamefont {Dabrowski}},\ }\bibfield
  {title} {\enquote {\bibinfo {title} {Negative refraction in
  ferromagnet-superconductor superlattices},}\ }\href@noop {} {\bibfield
  {journal} {\bibinfo  {journal} {Phys. Rev. Lett.}\ }\textbf {\bibinfo
  {volume} {95}},\ \bibinfo {pages} {247009} (\bibinfo {year}
  {2005})}\BibitemShut {NoStop}%
\bibitem [{\citenamefont {Haidar}\ \emph {et~al.}(2015)\citenamefont {Haidar},
  \citenamefont {Ranjbar}, \citenamefont {Balinsky}, \citenamefont {Dumas},
  \citenamefont {Khartsev},\ and\ \citenamefont
  {Akerman}}]{Haidar_JAP_117_17D119}%
  \BibitemOpen
  \bibfield  {author} {\bibinfo {author} {\bibfnamefont {M.}~\bibnamefont
  {Haidar}}, \bibinfo {author} {\bibfnamefont {M.}~\bibnamefont {Ranjbar}},
  \bibinfo {author} {\bibfnamefont {M.}~\bibnamefont {Balinsky}}, \bibinfo
  {author} {\bibfnamefont {R.~K.}\ \bibnamefont {Dumas}}, \bibinfo {author}
  {\bibfnamefont {S.}~\bibnamefont {Khartsev}}, \ and\ \bibinfo {author}
  {\bibfnamefont {J.}~\bibnamefont {Akerman}},\ }\bibfield  {title} {\enquote
  {\bibinfo {title} {Thickness- and temperature-dependent magnetodynamic
  properties of yttrium iron garnet thin films},}\ }\href@noop {} {\bibfield
  {journal} {\bibinfo  {journal} {J. Appl. Phys.}\ }\textbf {\bibinfo {volume}
  {117}},\ \bibinfo {pages} {17D119} (\bibinfo {year} {2015})}\BibitemShut
  {NoStop}%
\bibitem [{\citenamefont {Beaulieu}\ \emph {et~al.}(2018)\citenamefont
  {Beaulieu}, \citenamefont {Kervarec}, \citenamefont {Thiery}, \citenamefont
  {Klein}, \citenamefont {Naletov}, \citenamefont {Hurdequint}, \citenamefont
  {de~Loubens}, \citenamefont {Youssef},\ and\ \citenamefont
  {Vukadinovic}}]{Beaulieu_IEEEMagLett_9_3706005}%
  \BibitemOpen
  \bibfield  {author} {\bibinfo {author} {\bibfnamefont {N.}~\bibnamefont
  {Beaulieu}}, \bibinfo {author} {\bibfnamefont {N.}~\bibnamefont {Kervarec}},
  \bibinfo {author} {\bibfnamefont {N.}~\bibnamefont {Thiery}}, \bibinfo
  {author} {\bibfnamefont {O.}~\bibnamefont {Klein}}, \bibinfo {author}
  {\bibfnamefont {V.}~\bibnamefont {Naletov}}, \bibinfo {author} {\bibfnamefont
  {H.}~\bibnamefont {Hurdequint}}, \bibinfo {author} {\bibfnamefont
  {G.}~\bibnamefont {de~Loubens}}, \bibinfo {author} {\bibfnamefont {J.~B.}\
  \bibnamefont {Youssef}}, \ and\ \bibinfo {author} {\bibfnamefont
  {N.}~\bibnamefont {Vukadinovic}},\ }\bibfield  {title} {\enquote {\bibinfo
  {title} {Temperature dependence of magnetic properties of a ultrathin
  yttrium-iron garnet film grown by liquid phase epitaxy: Effect of a {Pt}
  overlayer},}\ }\href@noop {} {\bibfield  {journal} {\bibinfo  {journal} {IEEE
  Magnetics Letters}\ }\textbf {\bibinfo {volume} {9}},\ \bibinfo {pages}
  {3706005} (\bibinfo {year} {2018})}\BibitemShut {NoStop}%
\bibitem [{\citenamefont {Boventer}\ \emph {et~al.}(2018)\citenamefont
  {Boventer}, \citenamefont {Pfirrmann}, \citenamefont {Krause}, \citenamefont
  {Sch{\"o}n}, \citenamefont {Kl{\"a}ui},\ and\ \citenamefont
  {Weides}}]{Boventer_PRB_97_184420}%
  \BibitemOpen
  \bibfield  {author} {\bibinfo {author} {\bibfnamefont {I.}~\bibnamefont
  {Boventer}}, \bibinfo {author} {\bibfnamefont {M.}~\bibnamefont {Pfirrmann}},
  \bibinfo {author} {\bibfnamefont {J.}~\bibnamefont {Krause}}, \bibinfo
  {author} {\bibfnamefont {Y.}~\bibnamefont {Sch{\"o}n}}, \bibinfo {author}
  {\bibfnamefont {M.}~\bibnamefont {Kl{\"a}ui}}, \ and\ \bibinfo {author}
  {\bibfnamefont {M.}~\bibnamefont {Weides}},\ }\bibfield  {title} {\enquote
  {\bibinfo {title} {Complex temperature dependence of coupling and dissipation
  of cavity magnon polaritons from millikelvin to room temperature},}\
  }\href@noop {} {\bibfield  {journal} {\bibinfo  {journal} {Phys. Rev. B}\
  }\textbf {\bibinfo {volume} {97}},\ \bibinfo {pages} {184420} (\bibinfo
  {year} {2018})}\BibitemShut {NoStop}%
\bibitem [{\citenamefont {Neudecker}\ \emph {et~al.}(2006)\citenamefont
  {Neudecker}, \citenamefont {Woltersdorf}, \citenamefont {Heinrich},
  \citenamefont {Okuno}, \citenamefont {Gubbiotti},\ and\ \citenamefont
  {Back}}]{Neudecker_JMMM_307_148}%
  \BibitemOpen
  \bibfield  {author} {\bibinfo {author} {\bibfnamefont {I.}~\bibnamefont
  {Neudecker}}, \bibinfo {author} {\bibfnamefont {G.}~\bibnamefont
  {Woltersdorf}}, \bibinfo {author} {\bibfnamefont {B.}~\bibnamefont
  {Heinrich}}, \bibinfo {author} {\bibfnamefont {T.}~\bibnamefont {Okuno}},
  \bibinfo {author} {\bibfnamefont {G.}~\bibnamefont {Gubbiotti}}, \ and\
  \bibinfo {author} {\bibfnamefont {C.H.}\ \bibnamefont {Back}},\ }\bibfield
  {title} {\enquote {\bibinfo {title} {Comparison of frequency, field, and time
  domain ferromagnetic resonance methods},}\ }\href@noop {} {\bibfield
  {journal} {\bibinfo  {journal} {J. Magn. Magn. Mat.}\ }\textbf {\bibinfo
  {volume} {307}},\ \bibinfo {pages} {148} (\bibinfo {year}
  {2006})}\BibitemShut {NoStop}%
\bibitem [{\citenamefont {Kalarickal}\ \emph {et~al.}(2006)\citenamefont
  {Kalarickal}, \citenamefont {Krivosik}, \citenamefont {Wu}, \citenamefont
  {Patton}, \citenamefont {Schneider}, \citenamefont {Kabos}, \citenamefont
  {Silva},\ and\ \citenamefont {Nibarger}}]{Kalarickal_JAP_99_093909}%
  \BibitemOpen
  \bibfield  {author} {\bibinfo {author} {\bibfnamefont {S.~S.}\ \bibnamefont
  {Kalarickal}}, \bibinfo {author} {\bibfnamefont {P.}~\bibnamefont
  {Krivosik}}, \bibinfo {author} {\bibfnamefont {M.}~\bibnamefont {Wu}},
  \bibinfo {author} {\bibfnamefont {C.~E.}\ \bibnamefont {Patton}}, \bibinfo
  {author} {\bibfnamefont {M.~L.}\ \bibnamefont {Schneider}}, \bibinfo {author}
  {\bibfnamefont {P.}~\bibnamefont {Kabos}}, \bibinfo {author} {\bibfnamefont
  {T.~J.}\ \bibnamefont {Silva}}, \ and\ \bibinfo {author} {\bibfnamefont
  {J.~P.}\ \bibnamefont {Nibarger}},\ }\bibfield  {title} {\enquote {\bibinfo
  {title} {Ferromagnetic resonance linewidth in metallic thin films:
  {C}omparison of measurement methods},}\ }\href@noop {} {\bibfield  {journal}
  {\bibinfo  {journal} {J. Appl. Phys.}\ }\textbf {\bibinfo {volume} {99}},\
  \bibinfo {pages} {093909} (\bibinfo {year} {2006})}\BibitemShut {NoStop}%
\bibitem [{\citenamefont {Chen}\ \emph {et~al.}(2007)\citenamefont {Chen},
  \citenamefont {Hung}, \citenamefont {Yao}, \citenamefont {Lee}, \citenamefont
  {Ji},\ and\ \citenamefont {Yu}}]{Chen_JAP_101_09C104}%
  \BibitemOpen
  \bibfield  {author} {\bibinfo {author} {\bibfnamefont {Y.-C.}\ \bibnamefont
  {Chen}}, \bibinfo {author} {\bibfnamefont {D.-S.}\ \bibnamefont {Hung}},
  \bibinfo {author} {\bibfnamefont {Y.-D.}\ \bibnamefont {Yao}}, \bibinfo
  {author} {\bibfnamefont {S.-F.}\ \bibnamefont {Lee}}, \bibinfo {author}
  {\bibfnamefont {H.-P.}\ \bibnamefont {Ji}}, \ and\ \bibinfo {author}
  {\bibfnamefont {C.}~\bibnamefont {Yu}},\ }\bibfield  {title} {\enquote
  {\bibinfo {title} {Ferromagnetic resonance study of thickness-dependent
  magnetization precession in {N}i80{F}e20 films},}\ }\href@noop {} {\bibfield
  {journal} {\bibinfo  {journal} {J. Appl. Phys.}\ }\textbf {\bibinfo {volume}
  {101}},\ \bibinfo {pages} {09C104} (\bibinfo {year} {2007})}\BibitemShut
  {NoStop}%
\bibitem [{\citenamefont {Dubs}\ \emph {et~al.}(2017)\citenamefont {Dubs},
  \citenamefont {Surzhenko}, \citenamefont {Linke}, \citenamefont {Danilewsky},
  \citenamefont {Br{\"u}ckner},\ and\ \citenamefont
  {Dellith}}]{Dubs_JPDAP_50_204005}%
  \BibitemOpen
  \bibfield  {author} {\bibinfo {author} {\bibfnamefont {C.}~\bibnamefont
  {Dubs}}, \bibinfo {author} {\bibfnamefont {O.}~\bibnamefont {Surzhenko}},
  \bibinfo {author} {\bibfnamefont {R.}~\bibnamefont {Linke}}, \bibinfo
  {author} {\bibfnamefont {A.}~\bibnamefont {Danilewsky}}, \bibinfo {author}
  {\bibfnamefont {U.}~\bibnamefont {Br{\"u}ckner}}, \ and\ \bibinfo {author}
  {\bibfnamefont {Jan}\ \bibnamefont {Dellith}},\ }\bibfield  {title} {\enquote
  {\bibinfo {title} {Sub-micrometer yttrium iron garnet {LPE} films with low
  ferromagnetic resonance losses},}\ }\href@noop {} {\bibfield  {journal}
  {\bibinfo  {journal} {J. Phys. D: Appl. Phys.}\ }\textbf {\bibinfo {volume}
  {50}},\ \bibinfo {pages} {204005} (\bibinfo {year} {2017})}\BibitemShut
  {NoStop}%
\bibitem [{\citenamefont {Golovchanskiy}\ \emph
  {et~al.}(2016{\natexlab{a}})\citenamefont {Golovchanskiy}, \citenamefont
  {Bolginov}, \citenamefont {Abramov}, \citenamefont {Stolyarov}, \citenamefont
  {Hamida}, \citenamefont {Chichkov}, \citenamefont {Roditchev},\ and\
  \citenamefont {Ryazanov}}]{Golovchanskiy_JAP}%
  \BibitemOpen
  \bibfield  {author} {\bibinfo {author} {\bibfnamefont {I.~A.}\ \bibnamefont
  {Golovchanskiy}}, \bibinfo {author} {\bibfnamefont {V.~V.}\ \bibnamefont
  {Bolginov}}, \bibinfo {author} {\bibfnamefont {N.~N.}\ \bibnamefont
  {Abramov}}, \bibinfo {author} {\bibfnamefont {V.~S.}\ \bibnamefont
  {Stolyarov}}, \bibinfo {author} {\bibfnamefont {A.~Ben}\ \bibnamefont
  {Hamida}}, \bibinfo {author} {\bibfnamefont {V.~I.}\ \bibnamefont
  {Chichkov}}, \bibinfo {author} {\bibfnamefont {D.}~\bibnamefont {Roditchev}},
  \ and\ \bibinfo {author} {\bibfnamefont {V.~V.}\ \bibnamefont {Ryazanov}},\
  }\bibfield  {title} {\enquote {\bibinfo {title} {Magnetization dynamics in
  dilute {P}d$_{1-x}${F}e$_x$ thin films and patterned microstructures
  considered for superconducting electronics},}\ }\href@noop {} {\bibfield
  {journal} {\bibinfo  {journal} {J. Appl. Phys.}\ }\textbf {\bibinfo {volume}
  {120}},\ \bibinfo {pages} {163902} (\bibinfo {year}
  {2016}{\natexlab{a}})}\BibitemShut {NoStop}%
\bibitem [{\citenamefont {Kittel}(1948)}]{Kittel_PR_73_155}%
  \BibitemOpen
  \bibfield  {author} {\bibinfo {author} {\bibfnamefont {C.}~\bibnamefont
  {Kittel}},\ }\bibfield  {title} {\enquote {\bibinfo {title} {On the theory of
  ferromagnetic resonance absorption},}\ }\href@noop {} {\bibfield  {journal}
  {\bibinfo  {journal} {Phys. Rev}\ }\textbf {\bibinfo {volume} {73}},\
  \bibinfo {pages} {155} (\bibinfo {year} {1948})}\BibitemShut {NoStop}%
\bibitem [{\citenamefont {Khivintsev}\ \emph {et~al.}(2010)\citenamefont
  {Khivintsev}, \citenamefont {Reisman}, \citenamefont {Lovejoy}, \citenamefont
  {Adam}, \citenamefont {Schneider}, \citenamefont {Camley},\ and\
  \citenamefont {Celinski}}]{Khivintsev_JAP_108_023907}%
  \BibitemOpen
  \bibfield  {author} {\bibinfo {author} {\bibfnamefont {Y.~V.}\ \bibnamefont
  {Khivintsev}}, \bibinfo {author} {\bibfnamefont {L.}~\bibnamefont {Reisman}},
  \bibinfo {author} {\bibfnamefont {J.}~\bibnamefont {Lovejoy}}, \bibinfo
  {author} {\bibfnamefont {R.}~\bibnamefont {Adam}}, \bibinfo {author}
  {\bibfnamefont {C.~M.}\ \bibnamefont {Schneider}}, \bibinfo {author}
  {\bibfnamefont {R.~E.}\ \bibnamefont {Camley}}, \ and\ \bibinfo {author}
  {\bibfnamefont {Z.~J.}\ \bibnamefont {Celinski}},\ }\bibfield  {title}
  {\enquote {\bibinfo {title} {Spin wave resonance excitation in ferromagnetic
  films using planar waveguide structures},}\ }\href@noop {} {\bibfield
  {journal} {\bibinfo  {journal} {J. Appl. Phys.}\ }\textbf {\bibinfo {volume}
  {108}},\ \bibinfo {pages} {023907} (\bibinfo {year} {2010})}\BibitemShut
  {NoStop}%
\bibitem [{\citenamefont {Klingler}\ \emph {et~al.}(2015)\citenamefont
  {Klingler}, \citenamefont {Chumak}, \citenamefont {Mewes}, \citenamefont
  {Khodadadi}, \citenamefont {Mewes}, \citenamefont {Dubs}, \citenamefont
  {Surzhenko}, \citenamefont {Hillebrands},\ and\ \citenamefont
  {Conca}}]{Klingler_JPDAP_48_015001}%
  \BibitemOpen
  \bibfield  {author} {\bibinfo {author} {\bibfnamefont {S.}~\bibnamefont
  {Klingler}}, \bibinfo {author} {\bibfnamefont {A.~V.}\ \bibnamefont
  {Chumak}}, \bibinfo {author} {\bibfnamefont {T.}~\bibnamefont {Mewes}},
  \bibinfo {author} {\bibfnamefont {B.}~\bibnamefont {Khodadadi}}, \bibinfo
  {author} {\bibfnamefont {C.}~\bibnamefont {Mewes}}, \bibinfo {author}
  {\bibfnamefont {C.}~\bibnamefont {Dubs}}, \bibinfo {author} {\bibfnamefont
  {O.}~\bibnamefont {Surzhenko}}, \bibinfo {author} {\bibfnamefont
  {B.}~\bibnamefont {Hillebrands}}, \ and\ \bibinfo {author} {\bibfnamefont
  {A.}~\bibnamefont {Conca}},\ }\bibfield  {title} {\enquote {\bibinfo {title}
  {Measurements of the exchange stiffness of {YIG} films using broadband
  ferromagnetic resonance techniques},}\ }\href@noop {} {\bibfield  {journal}
  {\bibinfo  {journal} {J. Phys. D: Appl. Phys.}\ }\textbf {\bibinfo {volume}
  {48}},\ \bibinfo {pages} {015001} (\bibinfo {year} {2015})}\BibitemShut
  {NoStop}%
\bibitem [{\citenamefont {Artman}\ and\ \citenamefont
  {Charap}(1978)}]{Artman_JAP_49_1587}%
  \BibitemOpen
  \bibfield  {author} {\bibinfo {author} {\bibfnamefont {J.~O.}\ \bibnamefont
  {Artman}}\ and\ \bibinfo {author} {\bibfnamefont {S.~H.}\ \bibnamefont
  {Charap}},\ }\bibfield  {title} {\enquote {\bibinfo {title} {Ferromagnetic
  resonance in periodic domain structures},}\ }\href@noop {} {\bibfield
  {journal} {\bibinfo  {journal} {J. Appl. Phys.}\ }\textbf {\bibinfo {volume}
  {49}},\ \bibinfo {pages} {1587} (\bibinfo {year} {1978})}\BibitemShut
  {NoStop}%
\bibitem [{\citenamefont {Ramesh}\ and\ \citenamefont
  {Wigen}(1988)}]{RAMESH_JMMM_74_123}%
  \BibitemOpen
  \bibfield  {author} {\bibinfo {author} {\bibfnamefont {M.}~\bibnamefont
  {Ramesh}}\ and\ \bibinfo {author} {\bibfnamefont {P.~E.}\ \bibnamefont
  {Wigen}},\ }\bibfield  {title} {\enquote {\bibinfo {title}
  {Ferromagnetodynamics of parallel stripe domains - domain walls system},}\
  }\href@noop {} {\bibfield  {journal} {\bibinfo  {journal} {J. Mag. Mag.
  Mater.}\ }\textbf {\bibinfo {volume} {74}},\ \bibinfo {pages} {123} (\bibinfo
  {year} {1988})}\BibitemShut {NoStop}%
\bibitem [{\citenamefont {Camara}\ \emph {et~al.}(2017)\citenamefont {Camara},
  \citenamefont {Tacchi}, \citenamefont {Garnier}, \citenamefont {Eddrief},
  \citenamefont {Fortuna}, \citenamefont {Carlotti},\ and\ \citenamefont
  {Marangolo}}]{Camara_JPCM_29_465803}%
  \BibitemOpen
  \bibfield  {author} {\bibinfo {author} {\bibfnamefont {I.~S.}\ \bibnamefont
  {Camara}}, \bibinfo {author} {\bibfnamefont {S.}~\bibnamefont {Tacchi}},
  \bibinfo {author} {\bibfnamefont {L-.~C.}\ \bibnamefont {Garnier}}, \bibinfo
  {author} {\bibfnamefont {M.}~\bibnamefont {Eddrief}}, \bibinfo {author}
  {\bibfnamefont {F.}~\bibnamefont {Fortuna}}, \bibinfo {author} {\bibfnamefont
  {G.}~\bibnamefont {Carlotti}}, \ and\ \bibinfo {author} {\bibfnamefont
  {M.}~\bibnamefont {Marangolo}},\ }\bibfield  {title} {\enquote {\bibinfo
  {title} {Magnetization dynamics of weak stripe domains in {Fe}-{N} thin
  films: a multi-technique complementary approach},}\ }\href@noop {} {\bibfield
   {journal} {\bibinfo  {journal} {J. Phys.: Condens. Matter}\ }\textbf
  {\bibinfo {volume} {29}},\ \bibinfo {pages} {465803} (\bibinfo {year}
  {2017})}\BibitemShut {NoStop}%
\bibitem [{\citenamefont {Blake}\ \emph {et~al.}(1982)\citenamefont {Blake},
  \citenamefont {Shir}, \citenamefont {Tu},\ and\ \citenamefont
  {Torre}}]{BLAKE_IEEE_18_985}%
  \BibitemOpen
  \bibfield  {author} {\bibinfo {author} {\bibfnamefont {T.~G.~W.}\
  \bibnamefont {Blake}}, \bibinfo {author} {\bibfnamefont {C.-C.}\ \bibnamefont
  {Shir}}, \bibinfo {author} {\bibfnamefont {Y.-0}\ \bibnamefont {Tu}}, \ and\
  \bibinfo {author} {\bibfnamefont {E.~D.}\ \bibnamefont {Torre}},\ }\bibfield
  {title} {\enquote {\bibinfo {title} {Effects of finite anisotropy parameter
  {Q} in the determination of magnetic bubble material parameters},}\
  }\href@noop {} {\bibfield  {journal} {\bibinfo  {journal} {IEEE Trans.
  Magn.}\ }\textbf {\bibinfo {volume} {18}},\ \bibinfo {pages} {985} (\bibinfo
  {year} {1982})}\BibitemShut {NoStop}%
\bibitem [{\citenamefont {Virot}\ \emph {et~al.}(2012)\citenamefont {Virot},
  \citenamefont {Favre}, \citenamefont {Hayn},\ and\ \citenamefont
  {Kuz'min}}]{Virot_JPDAP_45_405003}%
  \BibitemOpen
  \bibfield  {author} {\bibinfo {author} {\bibfnamefont {F.}~\bibnamefont
  {Virot}}, \bibinfo {author} {\bibfnamefont {L.}~\bibnamefont {Favre}},
  \bibinfo {author} {\bibfnamefont {R.}~\bibnamefont {Hayn}}, \ and\ \bibinfo
  {author} {\bibfnamefont {M.~D.}\ \bibnamefont {Kuz'min}},\ }\bibfield
  {title} {\enquote {\bibinfo {title} {Theory of magnetic domains in uniaxial
  thin films},}\ }\href@noop {} {\bibfield  {journal} {\bibinfo  {journal} {J.
  Phys. D: Appl. Phys.}\ }\textbf {\bibinfo {volume} {45}},\ \bibinfo {pages}
  {405003} (\bibinfo {year} {2012})}\BibitemShut {NoStop}%
\bibitem [{\citenamefont {Lee}\ \emph {et~al.}(2016)\citenamefont {Lee},
  \citenamefont {Grudichak}, \citenamefont {Sklenar}, \citenamefont {Tsai},
  \citenamefont {Jang}, \citenamefont {Yang}, \citenamefont {Zhang},\ and\
  \citenamefont {Ketterson}}]{Lee_JAP_120_033905}%
  \BibitemOpen
  \bibfield  {author} {\bibinfo {author} {\bibfnamefont {S.}~\bibnamefont
  {Lee}}, \bibinfo {author} {\bibfnamefont {S.}~\bibnamefont {Grudichak}},
  \bibinfo {author} {\bibfnamefont {J.}~\bibnamefont {Sklenar}}, \bibinfo
  {author} {\bibfnamefont {C.~C.}\ \bibnamefont {Tsai}}, \bibinfo {author}
  {\bibfnamefont {M.}~\bibnamefont {Jang}}, \bibinfo {author} {\bibfnamefont
  {Q.}~\bibnamefont {Yang}}, \bibinfo {author} {\bibfnamefont {H.}~\bibnamefont
  {Zhang}}, \ and\ \bibinfo {author} {\bibfnamefont {J.~B.}\ \bibnamefont
  {Ketterson}},\ }\bibfield  {title} {\enquote {\bibinfo {title} {Ferromagnetic
  resonance of a {YIG} film in the low frequency regime},}\ }\href@noop {}
  {\bibfield  {journal} {\bibinfo  {journal} {J. Appl. Phys.}\ }\textbf
  {\bibinfo {volume} {120}},\ \bibinfo {pages} {033905} (\bibinfo {year}
  {2016})}\BibitemShut {NoStop}%
\bibitem [{\citenamefont {Manuilov}\ \emph {et~al.}(2009)\citenamefont
  {Manuilov}, \citenamefont {Khartsev},\ and\ \citenamefont
  {Grishin}}]{Manuilov_JAP_106_123917}%
  \BibitemOpen
  \bibfield  {author} {\bibinfo {author} {\bibfnamefont {S.~A.}\ \bibnamefont
  {Manuilov}}, \bibinfo {author} {\bibfnamefont {S.~I.}\ \bibnamefont
  {Khartsev}}, \ and\ \bibinfo {author} {\bibfnamefont {A.~M.}\ \bibnamefont
  {Grishin}},\ }\bibfield  {title} {\enquote {\bibinfo {title} {Pulsed laser
  deposited {Y3Fe5O12} films: Nature of magnetic anisotropy {I}},}\ }\href@noop
  {} {\bibfield  {journal} {\bibinfo  {journal} {J. Appl. Phys.}\ }\textbf
  {\bibinfo {volume} {106}},\ \bibinfo {pages} {123917} (\bibinfo {year}
  {2009})}\BibitemShut {NoStop}%
\bibitem [{\citenamefont {Stancil}(1993)}]{Stancil}%
  \BibitemOpen
  \bibfield  {author} {\bibinfo {author} {\bibfnamefont {D.}~\bibnamefont
  {Stancil}},\ }\href@noop {} {\emph {\bibinfo {title} {Theory of Magnetostatic
  Waves}}}\ (\bibinfo  {publisher} {Springer-Verlag New York, Inc.},\ \bibinfo
  {year} {1993})\BibitemShut {NoStop}%
\bibitem [{\citenamefont {Serga}\ \emph {et~al.}(2010)\citenamefont {Serga},
  \citenamefont {Chumak},\ and\ \citenamefont
  {Hillebrands}}]{Serga_JPDAP_43_264002}%
  \BibitemOpen
  \bibfield  {author} {\bibinfo {author} {\bibfnamefont {A.~A.}\ \bibnamefont
  {Serga}}, \bibinfo {author} {\bibfnamefont {A.~V.}\ \bibnamefont {Chumak}}, \
  and\ \bibinfo {author} {\bibfnamefont {B.}~\bibnamefont {Hillebrands}},\
  }\bibfield  {title} {\enquote {\bibinfo {title} {{YIG} magnonics},}\
  }\href@noop {} {\bibfield  {journal} {\bibinfo  {journal} {J. Phys. D: Appl.
  Phys.}\ }\textbf {\bibinfo {volume} {43}},\ \bibinfo {pages} {264002}
  (\bibinfo {year} {2010})}\BibitemShut {NoStop}%
\bibitem [{\citenamefont {Kittel}(1958)}]{Kittel_PR_100_1295}%
  \BibitemOpen
  \bibfield  {author} {\bibinfo {author} {\bibfnamefont {C.}~\bibnamefont
  {Kittel}},\ }\bibfield  {title} {\enquote {\bibinfo {title} {Excitation of
  spin waves in a ferromagnet by a uniform rf field},}\ }\href@noop {}
  {\bibfield  {journal} {\bibinfo  {journal} {Phys. Rev.}\ }\textbf {\bibinfo
  {volume} {100}},\ \bibinfo {pages} {1295} (\bibinfo {year}
  {1958})}\BibitemShut {NoStop}%
\bibitem [{\citenamefont {Seavey}\ and\ \citenamefont
  {Tannenwald}(1959)}]{Seavey_JAP_30_S227}%
  \BibitemOpen
  \bibfield  {author} {\bibinfo {author} {\bibfnamefont {M.~H.}\ \bibnamefont
  {Seavey}}\ and\ \bibinfo {author} {\bibfnamefont {P.~E.}\ \bibnamefont
  {Tannenwald}},\ }\bibfield  {title} {\enquote {\bibinfo {title} {Direct
  observation of spin wave resonance},}\ }\href@noop {} {\bibfield  {journal}
  {\bibinfo  {journal} {J. Appl. Phys.}\ }\textbf {\bibinfo {volume} {30}},\
  \bibinfo {pages} {S227} (\bibinfo {year} {1959})}\BibitemShut {NoStop}%
\bibitem [{\citenamefont {Bunyaev}\ \emph {et~al.}(2015)\citenamefont
  {Bunyaev}, \citenamefont {Golub}, \citenamefont {Salyuk}, \citenamefont
  {Tartakovskaya}, \citenamefont {Santos}, \citenamefont {Timopheev},
  \citenamefont {Sobolev}, \citenamefont {Serga}, \citenamefont {Chumak},
  \citenamefont {Hillebrands},\ and\ \citenamefont
  {Kakazei}}]{Bunyaev_SRep_5_18480}%
  \BibitemOpen
  \bibfield  {author} {\bibinfo {author} {\bibfnamefont {S.~A.}\ \bibnamefont
  {Bunyaev}}, \bibinfo {author} {\bibfnamefont {V.~O.}\ \bibnamefont {Golub}},
  \bibinfo {author} {\bibfnamefont {O.~Yu.}\ \bibnamefont {Salyuk}}, \bibinfo
  {author} {\bibfnamefont {E.~V.}\ \bibnamefont {Tartakovskaya}}, \bibinfo
  {author} {\bibfnamefont {N.~M.}\ \bibnamefont {Santos}}, \bibinfo {author}
  {\bibfnamefont {A.~A.}\ \bibnamefont {Timopheev}}, \bibinfo {author}
  {\bibfnamefont {N.~A.}\ \bibnamefont {Sobolev}}, \bibinfo {author}
  {\bibfnamefont {A.~A.}\ \bibnamefont {Serga}}, \bibinfo {author}
  {\bibfnamefont {A.~V.}\ \bibnamefont {Chumak}}, \bibinfo {author}
  {\bibfnamefont {B.}~\bibnamefont {Hillebrands}}, \ and\ \bibinfo {author}
  {\bibfnamefont {G.~N.}\ \bibnamefont {Kakazei}},\ }\bibfield  {title}
  {\enquote {\bibinfo {title} {Splitting of standing spin-wave modes in
  circular submicron ferromagnetic dot under axial symmetry violation},}\
  }\href@noop {} {\bibfield  {journal} {\bibinfo  {journal} {Sci. Rep.}\
  }\textbf {\bibinfo {volume} {5}},\ \bibinfo {pages} {18480} (\bibinfo {year}
  {2015})}\BibitemShut {NoStop}%
\bibitem [{\citenamefont {Smit}\ and\ \citenamefont
  {Beljers}(1955)}]{SMIT_PRR_10_113}%
  \BibitemOpen
  \bibfield  {author} {\bibinfo {author} {\bibfnamefont {J.}~\bibnamefont
  {Smit}}\ and\ \bibinfo {author} {\bibfnamefont {H.~G.}\ \bibnamefont
  {Beljers}},\ }\bibfield  {title} {\enquote {\bibinfo {title} {Ferromagnetic
  resonance absorption in {BaFe}$_{12}${O}$_{19}$ highly anisotropic
  crystal},}\ }\href@noop {} {\bibfield  {journal} {\bibinfo  {journal}
  {Philips Res. Rep.}\ }\textbf {\bibinfo {volume} {10}},\ \bibinfo {pages}
  {113} (\bibinfo {year} {1955})}\BibitemShut {NoStop}%
\bibitem [{\citenamefont {Suhl}(1955)}]{SUHL_PR_97_555}%
  \BibitemOpen
  \bibfield  {author} {\bibinfo {author} {\bibfnamefont {H.}~\bibnamefont
  {Suhl}},\ }\bibfield  {title} {\enquote {\bibinfo {title} {Ferromagnetic
  resonance in nickel ferrite between one and two kilomegacycles},}\
  }\href@noop {} {\bibfield  {journal} {\bibinfo  {journal} {Phys. Rev.}\
  }\textbf {\bibinfo {volume} {97}},\ \bibinfo {pages} {555} (\bibinfo {year}
  {1955})}\BibitemShut {NoStop}%
\bibitem [{\citenamefont {Rezende}\ \emph {et~al.}(1994)\citenamefont
  {Rezende}, \citenamefont {Moura}, \citenamefont {de~Aguiar},\ and\
  \citenamefont {Schreiner}}]{Rezende_PRB_49_15105}%
  \BibitemOpen
  \bibfield  {author} {\bibinfo {author} {\bibfnamefont {S.~M.}\ \bibnamefont
  {Rezende}}, \bibinfo {author} {\bibfnamefont {J.~A.~S.}\ \bibnamefont
  {Moura}}, \bibinfo {author} {\bibfnamefont {F.~M.}\ \bibnamefont
  {de~Aguiar}}, \ and\ \bibinfo {author} {\bibfnamefont {W.~H.}\ \bibnamefont
  {Schreiner}},\ }\bibfield  {title} {\enquote {\bibinfo {title} {Ferromagnetic
  resonance of {Fe}(111) thin films and {Fe}(111)/{Cu}(111) multilayers},}\
  }\href@noop {} {\bibfield  {journal} {\bibinfo  {journal} {Phys. Rev. B}\
  }\textbf {\bibinfo {volume} {49}},\ \bibinfo {pages} {15105} (\bibinfo {year}
  {1994})}\BibitemShut {NoStop}%
\bibitem [{\citenamefont {Hansen}(1974)}]{Hansen_JAP_45_3638}%
  \BibitemOpen
  \bibfield  {author} {\bibinfo {author} {\bibfnamefont {P.}~\bibnamefont
  {Hansen}},\ }\bibfield  {title} {\enquote {\bibinfo {title} {Anisotropy and
  magnetostriction of gallium‐substituted yttrium iron garnet},}\ }\href@noop
  {} {\bibfield  {journal} {\bibinfo  {journal} {J. Appl. Phys.}\ }\textbf
  {\bibinfo {volume} {45}},\ \bibinfo {pages} {3638} (\bibinfo {year}
  {1974})}\BibitemShut {NoStop}%
\bibitem [{\citenamefont {Howe}\ \emph {et~al.}(2015)\citenamefont {Howe},
  \citenamefont {Emori}, \citenamefont {Jeon}, \citenamefont {Oxholm},
  \citenamefont {Jones}, \citenamefont {Mahalingam}, \citenamefont {Zhuang},
  \citenamefont {Sun},\ and\ \citenamefont
  {Brown}}]{Howe_IEEEMagLet_6_3500504}%
  \BibitemOpen
  \bibfield  {author} {\bibinfo {author} {\bibfnamefont {Brandon~M.}\
  \bibnamefont {Howe}}, \bibinfo {author} {\bibfnamefont {Satoru}\ \bibnamefont
  {Emori}}, \bibinfo {author} {\bibfnamefont {Hyung-Min}\ \bibnamefont {Jeon}},
  \bibinfo {author} {\bibfnamefont {Trevor~M.}\ \bibnamefont {Oxholm}},
  \bibinfo {author} {\bibfnamefont {John~G.}\ \bibnamefont {Jones}}, \bibinfo
  {author} {\bibfnamefont {Krishnamurthy}\ \bibnamefont {Mahalingam}}, \bibinfo
  {author} {\bibfnamefont {Yan}\ \bibnamefont {Zhuang}}, \bibinfo {author}
  {\bibfnamefont {Nian~X.}\ \bibnamefont {Sun}}, \ and\ \bibinfo {author}
  {\bibfnamefont {Gail~J.}\ \bibnamefont {Brown}},\ }\bibfield  {title}
  {\enquote {\bibinfo {title} {Pseudomorphic yttrium iron garnet thin films
  with low damping and inhomogeneous linewidth broadening},}\ }\href@noop {}
  {\bibfield  {journal} {\bibinfo  {journal} {IEEE Magnetics Letters}\ }\textbf
  {\bibinfo {volume} {6}},\ \bibinfo {pages} {3500504} (\bibinfo {year}
  {2015})}\BibitemShut {NoStop}%
\bibitem [{\citenamefont {Bhoi}\ \emph {et~al.}(2018)\citenamefont {Bhoi},
  \citenamefont {Kim}, \citenamefont {Kim}, \citenamefont {Kim}, \citenamefont
  {Lee},\ and\ \citenamefont {Kim}}]{Bhoi_JAP_123_203902}%
  \BibitemOpen
  \bibfield  {author} {\bibinfo {author} {\bibfnamefont {Biswanath}\
  \bibnamefont {Bhoi}}, \bibinfo {author} {\bibfnamefont {Bosung}\ \bibnamefont
  {Kim}}, \bibinfo {author} {\bibfnamefont {Yongsub}\ \bibnamefont {Kim}},
  \bibinfo {author} {\bibfnamefont {Min-Kwan}\ \bibnamefont {Kim}}, \bibinfo
  {author} {\bibfnamefont {Jae-Hyeok}\ \bibnamefont {Lee}}, \ and\ \bibinfo
  {author} {\bibfnamefont {Sang-Koog}\ \bibnamefont {Kim}},\ }\bibfield
  {title} {\enquote {\bibinfo {title} {Stress-induced magnetic properties of
  {PLD}-grown high-quality ultrathin {YIG} films},}\ }\href@noop {} {\bibfield
  {journal} {\bibinfo  {journal} {J. Appl. Phys.}\ }\textbf {\bibinfo {volume}
  {123}},\ \bibinfo {pages} {203902} (\bibinfo {year} {2018})}\BibitemShut
  {NoStop}%
\bibitem [{\citenamefont {Maier-Flaig}\ \emph {et~al.}(2017)\citenamefont
  {Maier-Flaig}, \citenamefont {Klingler}, \citenamefont {Dubs}, \citenamefont
  {Surzhenko}, \citenamefont {Gross}, \citenamefont {Weiler}, \citenamefont
  {Huebl},\ and\ \citenamefont {Goennenwein}}]{Maier-Flaig_PRB_95_214423}%
  \BibitemOpen
  \bibfield  {author} {\bibinfo {author} {\bibfnamefont {H.}~\bibnamefont
  {Maier-Flaig}}, \bibinfo {author} {\bibfnamefont {S.}~\bibnamefont
  {Klingler}}, \bibinfo {author} {\bibfnamefont {C.}~\bibnamefont {Dubs}},
  \bibinfo {author} {\bibfnamefont {O.}~\bibnamefont {Surzhenko}}, \bibinfo
  {author} {\bibfnamefont {R.}~\bibnamefont {Gross}}, \bibinfo {author}
  {\bibfnamefont {M.}~\bibnamefont {Weiler}}, \bibinfo {author} {\bibfnamefont
  {H.}~\bibnamefont {Huebl}}, \ and\ \bibinfo {author} {\bibfnamefont
  {S.~T.~B.}\ \bibnamefont {Goennenwein}},\ }\bibfield  {title} {\enquote
  {\bibinfo {title} {Temperature-dependent magnetic damping of yttrium iron
  garnet spheres},}\ }\href@noop {} {\bibfield  {journal} {\bibinfo  {journal}
  {Phys. Rev. B}\ }\textbf {\bibinfo {volume} {95}},\ \bibinfo {pages} {214423}
  (\bibinfo {year} {2017})}\BibitemShut {NoStop}%
\bibitem [{\citenamefont {Jooss}\ \emph {et~al.}(2001)\citenamefont {Jooss},
  \citenamefont {Albrecht}, \citenamefont {Kuhn}, \citenamefont {Leonhardt},\
  and\ \citenamefont {Kronm{\"u}ller}}]{Jooss_RPP_65_651}%
  \BibitemOpen
  \bibfield  {author} {\bibinfo {author} {\bibfnamefont {Ch.}\ \bibnamefont
  {Jooss}}, \bibinfo {author} {\bibfnamefont {J.}~\bibnamefont {Albrecht}},
  \bibinfo {author} {\bibfnamefont {H.}~\bibnamefont {Kuhn}}, \bibinfo {author}
  {\bibfnamefont {S.}~\bibnamefont {Leonhardt}}, \ and\ \bibinfo {author}
  {\bibfnamefont {H.}~\bibnamefont {Kronm{\"u}ller}},\ }\bibfield  {title}
  {\enquote {\bibinfo {title} {Magneto-optical studies of current distributions
  in high-{Tc} superconductors},}\ }\href@noop {} {\bibfield  {journal}
  {\bibinfo  {journal} {Rep. Prog. Phys.}\ }\textbf {\bibinfo {volume} {65}},\
  \bibinfo {pages} {651} (\bibinfo {year} {2001})}\BibitemShut {NoStop}%
\bibitem [{\citenamefont {Wells}\ \emph {et~al.}(2016)\citenamefont {Wells},
  \citenamefont {Pan}, \citenamefont {Wilson}, \citenamefont {Golovchanskiy},
  \citenamefont {Fedoseev},\ and\ \citenamefont
  {Rozenfeld}}]{Wells_SUST_29_035014}%
  \BibitemOpen
  \bibfield  {author} {\bibinfo {author} {\bibfnamefont {F.~S.}\ \bibnamefont
  {Wells}}, \bibinfo {author} {\bibfnamefont {A.~V.}\ \bibnamefont {Pan}},
  \bibinfo {author} {\bibfnamefont {S.}~\bibnamefont {Wilson}}, \bibinfo
  {author} {\bibfnamefont {I.~A.}\ \bibnamefont {Golovchanskiy}}, \bibinfo
  {author} {\bibfnamefont {S.~A.}\ \bibnamefont {Fedoseev}}, \ and\ \bibinfo
  {author} {\bibfnamefont {A.}~\bibnamefont {Rozenfeld}},\ }\bibfield  {title}
  {\enquote {\bibinfo {title} {Dynamic magneto-optical imaging of
  superconducting thin films},}\ }\href@noop {} {\bibfield  {journal} {\bibinfo
   {journal} {Supercond. Sci. Technol.}\ }\textbf {\bibinfo {volume} {29}},\
  \bibinfo {pages} {035014} (\bibinfo {year} {2016})}\BibitemShut {NoStop}%
\bibitem [{\citenamefont {Wells}\ \emph {et~al.}(2017)\citenamefont {Wells},
  \citenamefont {Pan}, \citenamefont {Golovchanskiy}, \citenamefont
  {Fedoseev},\ and\ \citenamefont {Rozenfeld}}]{Wells_SRep_7_40235}%
  \BibitemOpen
  \bibfield  {author} {\bibinfo {author} {\bibfnamefont {F.~S.}\ \bibnamefont
  {Wells}}, \bibinfo {author} {\bibfnamefont {A.~V.}\ \bibnamefont {Pan}},
  \bibinfo {author} {\bibfnamefont {I.~A.}\ \bibnamefont {Golovchanskiy}},
  \bibinfo {author} {\bibfnamefont {S.~A.}\ \bibnamefont {Fedoseev}}, \ and\
  \bibinfo {author} {\bibfnamefont {A.}~\bibnamefont {Rozenfeld}},\ }\bibfield
  {title} {\enquote {\bibinfo {title} {Observation of transient overcritical
  currents in {YBCO} thin films using high-speed magneto-optical imaging and
  dynamic current mapping},}\ }\href@noop {} {\bibfield  {journal} {\bibinfo
  {journal} {Sci. Rep.}\ }\textbf {\bibinfo {volume} {7}},\ \bibinfo {pages}
  {40235} (\bibinfo {year} {2017})}\BibitemShut {NoStop}%
\bibitem [{\citenamefont {Bean}(1964)}]{BEAN_RMP_36_31}%
  \BibitemOpen
  \bibfield  {author} {\bibinfo {author} {\bibfnamefont {C.~P.}\ \bibnamefont
  {Bean}},\ }\bibfield  {title} {\enquote {\bibinfo {title} {Magnetization of
  high-field superconductors},}\ }\href@noop {} {\bibfield  {journal} {\bibinfo
   {journal} {Rev. Mod. Phys.}\ }\textbf {\bibinfo {volume} {36}},\ \bibinfo
  {pages} {31} (\bibinfo {year} {1964})}\BibitemShut {NoStop}%
\bibitem [{\citenamefont {Norris}(1969)}]{NORRIS_JPDAP_3_489}%
  \BibitemOpen
  \bibfield  {author} {\bibinfo {author} {\bibfnamefont {W.~T.}\ \bibnamefont
  {Norris}},\ }\bibfield  {title} {\enquote {\bibinfo {title} {Calculation of
  hysteresis losses in hard superconductors carrying ac: isolated conductors
  and edges of thin sheets},}\ }\href@noop {} {\bibfield  {journal} {\bibinfo
  {journal} {J. Phys. D: Appl. Phys.}\ }\textbf {\bibinfo {volume} {3}},\
  \bibinfo {pages} {489} (\bibinfo {year} {1969})}\BibitemShut {NoStop}%
\bibitem [{\citenamefont {Chen}\ and\ \citenamefont
  {Goldfarb}(1989)}]{Chen_JAP_66_2489}%
  \BibitemOpen
  \bibfield  {author} {\bibinfo {author} {\bibfnamefont {D.-X.}\ \bibnamefont
  {Chen}}\ and\ \bibinfo {author} {\bibfnamefont {R.~B.}\ \bibnamefont
  {Goldfarb}},\ }\bibfield  {title} {\enquote {\bibinfo {title} {Kim model for
  magnetization of type-{II} superconductors},}\ }\href@noop {} {\bibfield
  {journal} {\bibinfo  {journal} {J. Appl. Phys.}\ }\textbf {\bibinfo {volume}
  {66}},\ \bibinfo {pages} {2489} (\bibinfo {year} {1989})}\BibitemShut
  {NoStop}%
\bibitem [{\citenamefont {Golovchanskiy}\ \emph {et~al.}(2013)\citenamefont
  {Golovchanskiy}, \citenamefont {Pan}, \citenamefont {Shcherbakova},\ and\
  \citenamefont {Fedoseev}}]{Golovchanskiy_JAP_114_163910}%
  \BibitemOpen
  \bibfield  {author} {\bibinfo {author} {\bibfnamefont {I.~A.}\ \bibnamefont
  {Golovchanskiy}}, \bibinfo {author} {\bibfnamefont {A.~V.}\ \bibnamefont
  {Pan}}, \bibinfo {author} {\bibfnamefont {O.~V.}\ \bibnamefont
  {Shcherbakova}}, \ and\ \bibinfo {author} {\bibfnamefont {S.~A.}\
  \bibnamefont {Fedoseev}},\ }\bibfield  {title} {\enquote {\bibinfo {title}
  {Rectifying differences in transport, dynamic, and quasi-equilibrium
  measurements of critical current density},}\ }\href@noop {} {\bibfield
  {journal} {\bibinfo  {journal} {J. Appl. Phys.}\ }\textbf {\bibinfo {volume}
  {114}},\ \bibinfo {pages} {163910} (\bibinfo {year} {2013})}\BibitemShut
  {NoStop}%
\bibitem [{\citenamefont {Tinkham}(1988)}]{Tinkham_PRL_61_1658}%
  \BibitemOpen
  \bibfield  {author} {\bibinfo {author} {\bibfnamefont {M.}~\bibnamefont
  {Tinkham}},\ }\bibfield  {title} {\enquote {\bibinfo {title} {Resistive
  transition of high-temperature superconductors},}\ }\href@noop {} {\bibfield
  {journal} {\bibinfo  {journal} {Phys. Rev. Lett.}\ }\textbf {\bibinfo
  {volume} {61}},\ \bibinfo {pages} {1658} (\bibinfo {year}
  {1988})}\BibitemShut {NoStop}%
\bibitem [{\citenamefont {Golovchanskiy}\ \emph
  {et~al.}(2016{\natexlab{b}})\citenamefont {Golovchanskiy}, \citenamefont
  {Pan}, \citenamefont {George}, \citenamefont {Wells}, \citenamefont
  {Fedoseev},\ and\ \citenamefont {Rozenfeld}}]{Golovchanskiy_SUST_29_075002}%
  \BibitemOpen
  \bibfield  {author} {\bibinfo {author} {\bibfnamefont {I.~A.}\ \bibnamefont
  {Golovchanskiy}}, \bibinfo {author} {\bibfnamefont {A.~V.}\ \bibnamefont
  {Pan}}, \bibinfo {author} {\bibfnamefont {J.}~\bibnamefont {George}},
  \bibinfo {author} {\bibfnamefont {F.~S.}\ \bibnamefont {Wells}}, \bibinfo
  {author} {\bibfnamefont {S.~A.}\ \bibnamefont {Fedoseev}}, \ and\ \bibinfo
  {author} {\bibfnamefont {A.}~\bibnamefont {Rozenfeld}},\ }\bibfield  {title}
  {\enquote {\bibinfo {title} {Vibration effect on magnetization and critical
  current density of superconductors},}\ }\href@noop {} {\bibfield  {journal}
  {\bibinfo  {journal} {Supercond. Sci. Technol.}\ }\textbf {\bibinfo {volume}
  {29}},\ \bibinfo {pages} {075002} (\bibinfo {year}
  {2016}{\natexlab{b}})}\BibitemShut {NoStop}%
\bibitem [{\citenamefont {Jeon}\ \emph {et~al.}(2019)\citenamefont {Jeon},
  \citenamefont {Ciccarelli}, \citenamefont {Kurebayashi}, \citenamefont
  {Cohen}, \citenamefont {Montiel}, \citenamefont {Eschrig}, \citenamefont
  {Wagner}, \citenamefont {Komori}, \citenamefont {Srivastava}, \citenamefont
  {Robinson},\ and\ \citenamefont {Blamire}}]{Jeon_arXiv}%
  \BibitemOpen
  \bibfield  {author} {\bibinfo {author} {\bibfnamefont {K.}~\bibnamefont
  {Jeon}}, \bibinfo {author} {\bibfnamefont {C.}~\bibnamefont {Ciccarelli}},
  \bibinfo {author} {\bibfnamefont {H.}~\bibnamefont {Kurebayashi}}, \bibinfo
  {author} {\bibfnamefont {L.~F.}\ \bibnamefont {Cohen}}, \bibinfo {author}
  {\bibfnamefont {X.}~\bibnamefont {Montiel}}, \bibinfo {author} {\bibfnamefont
  {M.}~\bibnamefont {Eschrig}}, \bibinfo {author} {\bibfnamefont
  {T.}~\bibnamefont {Wagner}}, \bibinfo {author} {\bibfnamefont
  {S.}~\bibnamefont {Komori}}, \bibinfo {author} {\bibfnamefont
  {A.}~\bibnamefont {Srivastava}}, \bibinfo {author} {\bibfnamefont {J.~W.~A.}\
  \bibnamefont {Robinson}}, \ and\ \bibinfo {author} {\bibfnamefont {M.~G.}\
  \bibnamefont {Blamire}},\ }\bibfield  {title} {\enquote {\bibinfo {title}
  {Effect of meissner screening and trapped magnetic flux on magnetization
  dynamics in thick {Nb/Ni80Fe20/Nb} trilayers},}\ }\href@noop {} {\bibfield
  {journal} {\bibinfo  {journal} {Phys. Rev. Appl.}\ }\textbf {\bibinfo
  {volume} {11}},\ \bibinfo {pages} {014061} (\bibinfo {year}
  {2019})}\BibitemShut {NoStop}%
\bibitem [{\citenamefont {Gibbons}\ and\ \citenamefont
  {Chirba}(1958)}]{GIBBONS_PR_110_770}%
  \BibitemOpen
  \bibfield  {author} {\bibinfo {author} {\bibfnamefont {D.~F.}\ \bibnamefont
  {Gibbons}}\ and\ \bibinfo {author} {\bibfnamefont {V.~G.}\ \bibnamefont
  {Chirba}},\ }\bibfield  {title} {\enquote {\bibinfo {title} {Acoustical loss
  and young's modulus of yttrium iron garnet},}\ }\href@noop {} {\bibfield
  {journal} {\bibinfo  {journal} {Phys. Rev.}\ }\textbf {\bibinfo {volume}
  {110}},\ \bibinfo {pages} {770} (\bibinfo {year} {1958})}\BibitemShut
  {NoStop}%
\bibitem [{\citenamefont {Wang}\ and\ \citenamefont
  {Reeber}(1998)}]{Wang_MatSciEng_R23_101}%
  \BibitemOpen
  \bibfield  {author} {\bibinfo {author} {\bibfnamefont {Kai}\ \bibnamefont
  {Wang}}\ and\ \bibinfo {author} {\bibfnamefont {Robert~R.}\ \bibnamefont
  {Reeber}},\ }\bibfield  {title} {\enquote {\bibinfo {title} {The role of
  defects on thermophysical properties: Thermal expansion of {V}, {Nb}, {Ta},
  {Mo} and {W}},}\ }\href@noop {} {\bibfield  {journal} {\bibinfo  {journal}
  {Materials Science and Engineering}\ }\textbf {\bibinfo {volume} {R23}},\
  \bibinfo {pages} {101} (\bibinfo {year} {1998})}\BibitemShut {NoStop}%
\bibitem [{\citenamefont {Geller}\ \emph {et~al.}(1969)\citenamefont {Geller},
  \citenamefont {Espinosa},\ and\ \citenamefont
  {Crandall}}]{GELLER_Jr_Appl_Cryst_2_86}%
  \BibitemOpen
  \bibfield  {author} {\bibinfo {author} {\bibfnamefont {S.}~\bibnamefont
  {Geller}}, \bibinfo {author} {\bibfnamefont {G.~P.}\ \bibnamefont
  {Espinosa}}, \ and\ \bibinfo {author} {\bibfnamefont {P.~B.}\ \bibnamefont
  {Crandall}},\ }\bibfield  {title} {\enquote {\bibinfo {title} {Thermal
  expansion of yttrium and gadolinium iron, gallium and aluminum garnets},}\
  }\href@noop {} {\bibfield  {journal} {\bibinfo  {journal} {Jr. Appl. Cryst.}\
  }\textbf {\bibinfo {volume} {2}},\ \bibinfo {pages} {86} (\bibinfo {year}
  {1969})}\BibitemShut {NoStop}%
\bibitem [{\citenamefont {sheng Liang}\ and\ \citenamefont {chao
  Liu}(1999)}]{Liang_Powder_Diffraction_14_2}%
  \BibitemOpen
  \bibfield  {author} {\bibinfo {author} {\bibfnamefont {Rui}\ \bibnamefont
  {sheng Liang}}\ and\ \bibinfo {author} {\bibfnamefont {Feng}\ \bibnamefont
  {chao Liu}},\ }\bibfield  {title} {\enquote {\bibinfo {title} {Measurement of
  thermal expansion coeffiecient of substrate {GGG} and its epitaxial layer
  {YIG}},}\ }\href@noop {} {\bibfield  {journal} {\bibinfo  {journal} {Powder
  Diffraction}\ }\textbf {\bibinfo {volume} {14}},\ \bibinfo {pages} {2}
  (\bibinfo {year} {1999})}\BibitemShut {NoStop}%
\bibitem [{\citenamefont {Levinstein}\ \emph {et~al.}(1966)\citenamefont
  {Levinstein}, \citenamefont {Gyorgy},\ and\ \citenamefont
  {LeCraw}}]{Levinstein_JAP_37_2197}%
  \BibitemOpen
  \bibfield  {author} {\bibinfo {author} {\bibfnamefont {H.~J.}\ \bibnamefont
  {Levinstein}}, \bibinfo {author} {\bibfnamefont {E.~M.}\ \bibnamefont
  {Gyorgy}}, \ and\ \bibinfo {author} {\bibfnamefont {R.~C.}\ \bibnamefont
  {LeCraw}},\ }\bibfield  {title} {\enquote {\bibinfo {title} {Thermal
  expansion of {YIG} and {YIG} with {Mn} and {Si} additions},}\ }\href@noop {}
  {\bibfield  {journal} {\bibinfo  {journal} {J. Appl. Phys.}\ }\textbf
  {\bibinfo {volume} {37}},\ \bibinfo {pages} {2197} (\bibinfo {year}
  {1966})}\BibitemShut {NoStop}%
\bibitem [{\citenamefont {Antyukhov}\ \emph {et~al.}(1987)\citenamefont
  {Antyukhov}, \citenamefont {Sidorov}, \citenamefont {Ivanov},\ and\
  \citenamefont {Antonov}}]{Antyukhov_Inorg_Mater_23_702}%
  \BibitemOpen
  \bibfield  {author} {\bibinfo {author} {\bibfnamefont {A.~M.}\ \bibnamefont
  {Antyukhov}}, \bibinfo {author} {\bibfnamefont {A.~A.}\ \bibnamefont
  {Sidorov}}, \bibinfo {author} {\bibfnamefont {I.~A.}\ \bibnamefont {Ivanov}},
  \ and\ \bibinfo {author} {\bibfnamefont {A.~V.}\ \bibnamefont {Antonov}},\
  }\bibfield  {title} {\enquote {\bibinfo {title} {Thermal expansion
  coefficients of crystals of certain garnets over the range 6-310 {K}},}\
  }\href@noop {} {\bibfield  {journal} {\bibinfo  {journal} {Inorg. Mater.
  (Translated from Izv. Akad. Nauk SSSR, Neorg. Mater.)}\ }\textbf {\bibinfo
  {volume} {23}},\ \bibinfo {pages} {702} (\bibinfo {year} {1987})}\BibitemShut
  {NoStop}%
\bibitem [{\citenamefont {Fan}\ \emph {et~al.}(2007)\citenamefont {Fan},
  \citenamefont {Ripin}, \citenamefont {Aggarwal}, \citenamefont {Ochoa},
  \citenamefont {Chann}, \citenamefont {Tilleman},\ and\ \citenamefont
  {Spitzberg}}]{Fan_IEEE_13_448}%
  \BibitemOpen
  \bibfield  {author} {\bibinfo {author} {\bibfnamefont {Tso~Yee}\ \bibnamefont
  {Fan}}, \bibinfo {author} {\bibfnamefont {Daniel~J.}\ \bibnamefont {Ripin}},
  \bibinfo {author} {\bibfnamefont {Roshan~L.}\ \bibnamefont {Aggarwal}},
  \bibinfo {author} {\bibfnamefont {Juan~R.}\ \bibnamefont {Ochoa}}, \bibinfo
  {author} {\bibfnamefont {Bien}\ \bibnamefont {Chann}}, \bibinfo {author}
  {\bibfnamefont {Michael}\ \bibnamefont {Tilleman}}, \ and\ \bibinfo {author}
  {\bibfnamefont {Joshua}\ \bibnamefont {Spitzberg}},\ }\bibfield  {title}
  {\enquote {\bibinfo {title} {Cryogenic {Yb}$^{3+}$-doped solid-state
  lasers},}\ }\href@noop {} {\bibfield  {journal} {\bibinfo  {journal} {IEEE
  Journal of Selected Topics in Quantum Electronics}\ }\textbf {\bibinfo
  {volume} {13}},\ \bibinfo {pages} {448} (\bibinfo {year} {2007})}\BibitemShut
  {NoStop}%
\end{thebibliography}%

\end{document}